\documentclass[onecolumn,authoryear]{els-mrw} 

\def\aj{AJ}                      
\def\araa{ARA\&A}       
\def\apj{ApJ}                 
\def\apjl{ApJ}                
\def\aap{A\&A}              
\def\icarus{Icarus}         
\def\mnras{MNRAS}     
\def\nat{Nature}
\def\pasj{PASJ}

\usepackage{amsmath,amssymb,amsfonts,amsthm,makeidx,graphicx}
\usepackage{txfonts}
\usepackage{helvet}

\usepackage[dvipsnames]{xcolor}
\usepackage[colorlinks = true]{hyperref}

\usepackage{tocloft}
\renewcommand{\contentsname}{}

\begin{document}

\chapter*{Planet-disk interaction and evolution}\label{chap1}

\author[1]{Pablo Benítez-Llambay}%

\address[1]{\orgname{Universidad Adolfo Ib\'a\~nez}, \orgdiv{Facultad de Ingenier\'ia y Ciencias}, \orgaddress{Av. Diagonal las Torres 2640, Peñalol\'en, Chile.}}

\maketitle

\begin{abstract}[Abstract]
Following the groundbreaking discovery of the first extrasolar planet orbiting a sun-like star, 51 Pegasi b in 1995, the field of planet formation has become a cornerstone of modern astrophysics. This is in part due to the revelation of an astonishing diversity of planetary types and architectures, inferred from detailed astronomical observations. This diversity is driven by the interplay between the physical processes governing planet assembly and the environmental conditions within the protoplanetary disk in which they form. This chapter provides an introduction to the specific mechanisms that can induce orbital variations in nascent planetary bodies during their formation and evolution.
\end{abstract}

\begin{keywords}
Planet formation - circumstellar disks, circumstellar dust, exoplanets, planetary system formation, planetary system evolution.
\end{keywords}

\noindent \rule{\textwidth}{1.0pt} \\
\noindent {\Large \bf Contents}
\hfill \break 
\vspace{-3.3cm}
\tableofcontents
\vspace{0.25cm}
\noindent \rule{\textwidth}{1.0pt}

\addcontentsline{toc}{section}{Learning objectives}
\section*{Learning objectives}
By the end of this chapter, you should understand:
\\\\$\bullet$ \hspace{0.5cm} What is planet-disk interaction.
\\$\bullet$ \hspace{0.5cm} How is planet-disk interaction connected to planet migration.
\\$\bullet$ \hspace{0.5cm} How is planet-disk interaction modeled.
\\$\bullet$ \hspace{0.5cm} What are the most relevant parameters and quantities of the problem.
\\$\bullet$ \hspace{0.5cm} How planet-disk interaction behaves for different planetary masses and in different environments.
\\$\bullet$ \hspace{0.5cm} How could planets evolve because of planet-disk interaction?

\newpage

\begin{glossary}[Glossary]
\term{Barotropic Equation of State} An equation of state that relates pressure to density only.  \\
\term{Boundary Conditions} Conditions imposed on the edges of a physical system to define its behavior.  \\
\term{Corotation Torque} The torque exerted on a planet by the interaction of gas particles that are trapped or cross the horseshoe region around the planet. \\
\term{Convergent Migration} Relative migration of two or more planets approaching each other. \\
\term{Differential Lindblad Torque} The net torque exerted by disk spiral density waves at the inner and outer Lindblad resonances.  \\
\term{Disk Unperturbed Vortensity} The vortensity of the disk in the absence of a planet.  \\
\term{Epicyclic Frequency} The frequency at which a small radial disturbance in a disk oscillates.  \\
\term{Gap} Annular region of significantly reduced gas density in a disk. \\
\term{Gravitational Potential} The potential energy per unit mass at a given point in a gravitational field. \\
\term{Headwind} A wind that blows against the direction of motion of an object.  \\
\term{Heating Torque} A torque exerted on a planet due to the heating of the disk by the planet's luminosity.  \\
\term{Horseshoe Orbits} Orbits that librate around a planet without a net radial migration. \\
\term{Horseshoe semi-width} The half-width of the horseshoe region around a planet. \\
\term{Horseshoe Drag} The torque exerted on a planet by the motion of gas particles in the horseshoe region.  \\
\term{Interstellar Medium} The material that fills the space between stars, consisting of gas and dust.  \\
\term{Keplerian Angular Velocity} The angular velocity of a body in circular orbit around another body.  \\
\term{Lagrange Equilibrium Points} Points in the restricted three-body problem where a small object remain stationary in a corotating frame. \\
\term{Linear Corotation Torque} The corotation torque calculated using linear theory.  \\
\term{Linear Regime} The regime in which a system can be described by linear equations.  \\
\term{Magneto-Hydrodynamics} The study of the interaction of electrically conducting fluids with magnetic fields.  \\
\term{Magneto-Hydrodynamical Winds} Outflows driven by magnetic fields in a disk.  \\
\term{Mean-Motion Resonance} A configuration in which two or more planets orbit their star in such a way that their orbital periods are in a simple ratio.  \\
\term{Migration Rate} The rate at which a planet's orbital radius changes.  \\
\term{Molecular Cloud} A large cloud of gas and dust in interstellar space, primarily composed of hydrogen molecules.  \\
\term{Navier-Stokes Equations} A set of partial differential equations that describe the motion of viscous fluids.  \\
\term{Non-Linear Effects} Effects that cannot be described by linear equations.  \\
\term{Orbital Parameters} The characteristics of an object's orbit, including its shape, size, and orientation.  \\
\term{Planet-Disk Interaction} The gravitational and hydrodynamic forces between a forming planet and the protoplanetary disk.  \\
\term{Planetary Migration} The movement of a planet from its original position to a new location within the protoplanetary disk.  \\
\term{Plummer Potential} A softened gravitational potential used in two-dimensional simulations of planet-disk interaction.  \\
\term{Pressure Buffer} A mechanism that reduces the sensitivity of the differential Lindblad torque to the surface density gradient.  \\
\term{Protoplanetary Disk} A disk of gas and dust surrounding a young star from which planets can form.  \\
\term{Photoevaporation} The process by which gas in a disk is heated and escapes due to radiation from the central star or a nearby companion.  \\
\term{Radiation Hydrodynamics} The study of the interaction of fluids with radiation.  \\
\term{Resonant Angles} Angles that describe the phase difference between the orbits of two or more planets in a resonance.  \\
\term{Resonant Chain} A series of planets that are all in mean-motion resonances with their nearest neighbors.  \\
\term{Rossby Wave Instability} A linear instability that can lead to the formation of vortices.  \\
\term{Scale Height} The vertical distance over which the density of the disk decreases by a factor of $e$.  \\
\term{Shock} A discontinuity in the fluid variables.  \\
\term{Stagnation Points} Points in a fluid where the velocity is zero in the corotating frame.  \\
\term{Stokes Number} A dimensionless number that measures the degree of coupling between a particle and the gas in a disk. Large/small Stokes numbers are associated with large/small particles or low/high gas-density regions. \\
\term{Sub-Keplerian Motion} Motion at a speed that is less than the Keplerian speed.  \\
\term{Synodic Period} The time it takes for a celestial body to return to the same position relative to another celestial body.  \\
\term{Type-I migration} Migration of low-mass planets in laminar viscous disks. \\
\term{Type-II migration} Migration regime at which the planet migrates at the disk viscous speed.\\
\term{Type-III migration} Very fast migration of partial gap opening planets, typically sub-giants.\\
\term{Viscous Accretion} The process by which matter in a disk accretes onto a central star due to viscous forces.  \\
\term{Vortensity} the z-component of the fluid vorticity divided by the surface density. \\
\term{Vorticity} The curl of the velocity field. It measures the local rotation of a fluid element. \\
\end{glossary}

\newpage

\section*{List of Symbols}

\begin{table*}[h!]
\centering
\begin{tabular}[t]{p{1cm} p{0.33\linewidth}}
	Symbol & Definition \\
	\hline
	\hline
	$a$ & Semi-major axis of the planet's orbit \\
	$\alpha$ & Minus the logarithmic derivative of the gas surface density. \\
	$\alpha_\nu$ & Dimensionless viscosity parameter\\
	$\beta$ & Minus the logarithmic derivative of the gas temperature \\
	$c_{\rm s}$ & Sound speed\\
	$\gamma$ & Adiabatic index\\
	$\Gamma$ & z-component of the torque exerted by the disk \\
	$\Gamma_{\rm p}$ & Reference torque at the planet radius. \\
	$\Gamma^{\rm nmig}$ & Torque felt by a planet when it does not migrate\\
	$\Gamma^{\rm mig}$ & Additional torque felt by the planet when it migrates\\
	$e$ & Eccentricity of the planet's orbit \\
	$E_{\rm p}$ & Energy of the planet \\
	$\epsilon$ & Gas internal energy density \\
	$f_{\rm gap}$ & Fraction of the initial surface density at the bottom of a planet's gap. \\
	${\bf F}_{\rm d}$ & Gravitational force exerted by the disk on the planet \\
	$G$ & Gravitational constant \\
	$\eta$ & Dimensionless number that quantifies the sub-Keplerian motion of the gas \\
	$h$ & Disk aspect-ratio\\
	$H$ & Disk scale-height\\
	$\theta$ & Pitch angle\\
	${\bf I}$ & Identity matrix\\
	$j_{\rm p}$ & Angular momentum of the planet \\
	$\kappa$ & Epicyclic frequency \\
	$K$ & Dimensionless parameter of which the depth of a planet's gap depends on\\
	$l_{\rm s}$ & Distance from the planet at which the planet perturbation becomes as shock\\
	$m$ & Azimuthal wave number \\
	$m_{\rm d}$ & Vorticity weighted coorbital mass deficit\\
	$m_{\rm p}$ & Mass of the planet\\
	$M_\star$ & Mass of the star\\
	$m_{\rm th}$ & Thermal mass\\
	$\nu$ & Kinematic viscosity\\
	$\xi_0$ & Dimensionless strength of the torque\\
	\hline
\end{tabular}
\hspace{5em}
\begin{tabular}[t]{p{1cm} p{0.33\linewidth}}
	Symbol & Definition \\
	\hline
	\hline
	$\rho$ & Gas density\\
	$P$ & Gas pressure\\
	$\mathcal{P}$ & Energy per unit time exchanged between the disk and the planet \\
	$Q$ & Planet mass divided by the thermal mass\\
	$\bf{r}$ & Position vector \\
	$r$ & Cylindrical radius \\
	$r_{\rm L}$ & Radius of the Lindblad resonances \\
	$r_{\rm s}$ & Smoothing length of the planet's potential \\
	$\Sigma$ & Disk surface density \\
	$\Sigma_0$ & Unperturbed disk surface density at some fiducial radius\\
	$\Sigma_{\rm gap}$ & Disk surface density at the bottom of a planet's gap\\
	$\dot{\Sigma}_{\rm w}$ & Disk's mass loss rate due to winds\\
	$S_{\rm t}$ & Stokes number \\
	$t$ & Time\\
	$T$ & Viscous stress tensor \\
	$\tau_{\rm libration}$ & Libration time scale within the horseshoe region\\
	$\tau_a$ & Semi-major axis damping/excitation time scale or migration timescale\\
	$\tau_e$ & Eccentricity damping/excitation time scale\\
	$\tau_i$ & Inclination damping/excitation time scale\\
	$\tau_\nu$ & Viscous time scale \\
	$\tau_i$ & Inclination damping/excitation time scale\\
	$\varphi$ & Azimuthal angle \\
	$\varphi_{\rm p}$ & Azimuthal angle of the planet \\
	$\varphi_{\rm wake}$ & Azimuthal angle of the wake's peak density\\
	$\phi$ & Gravitational potential\\
	${\bf v}$ & Gas velocity vector\\
	${\bf v}_{\rm p}$ & Planet's velocity vector\\
	${\bf v}_{\rm solid}$ & Solid's velocity vector\\
	$x_{\rm s}$ & Semi-width of the horseshoe region\\
	$\psi$ & Dimensionless strength of the torque excerted by the disk onto the planet\\
	$\Omega$ & Angular velocity of the gas\\
	$\Omega_{\rm K}$ & Keplerian angular velocity\\
	$\Omega_{\rm p}$ & Angular velocity of the planet\\
	$\omega$ & Gas vortensity\\
	$z$ & Vertical cylindrical coordinate\\
	\hline
\end{tabular}
\end{table*}

\newpage 

While a comprehensive list of relevant works would be valuable, to avoid detracting from the primary educational objective of this chapter, we have been forced to focus on a select number of studies that provide a solid foundation for the topics presented. 
We encourage interested readers to complement our presentation with the reviews listed at the end of this chapter and references therein.

\section{Introduction - What is planet-disk interaction?}
A star is born after the gravitational collapse of a cold extended molecular cloud. As the material falls towards the protostar, angular momentum conservation and energy dissipation lead to the formation of a protoplanetary disk.
This disk shares the bulk composition of the interstellar medium, which is mainly made of hydrogen, a small fraction of heavier gasses, and micrometer-sized dust particles. 
The amount of gas and composition observed in giant planets, like Jupiter or Saturn in the Solar System, indicate that they grew embedded in a gaseous environment. Therefore, they should form embedded in the protoplanetary disk.
Protoplanetary disks disappear on $\mathcal{O}(10^6)-\mathcal{O}(10^7)$ yr, which sets an upper bound for giant planet formation timescale. 

A forming planet exerts a gravitational force onto the disk. Newton's third law implies that the disk must exert an equal (but opposite in direction) force onto the planet. These forces allow the exchange of momentum and energy between the planet and the disk, altering the disk structure which, in turn, can make the forming planet change its orbital parameters and move over large spatial scales ($\mathcal{O}(10)$ au) on relatively short timescales ($\mathcal{O}(10^5-10^6)$ yr). 
The fundamental mechanism is referred to as planet-disk interaction, resulting in planetary migration. This interaction can also impact the overall structure and evolution of the disk.
Characterizing the speed and direction of this migration requires a thorough understanding of the disk physical conditions both globally and in the vicinity of the forming planet. 

\section{Planet-Disk Interaction and Planet Migration}
\label{sec:planet-disk-migration}

To illustrate the effect of planet-disk interaction and the resulting migration, consider the simple model of a planet of mass $m_{\rm p}$ orbiting a central star of mass $M_\star$ in a coplanar elliptical orbit with the protoplanetary disk. The orbit is characterized by a semi-major axis, $a$, and eccentricity, $e$.  
The angular momentum, $j_{\rm p}$, and energy, $E_{\rm p}$, of the planet are 
\begin{align}
j_{\rm p} &= m_{\rm p} \sqrt{G M_\star a (1-e^2)}\,, \\
E_{\rm p} &= - \frac{G M_{\star} m_{\rm p} }{2a}\,,
\end{align}
where $G$ is the gravitational constant. The protoplanetary disk exerts a gravitational force, denoted by ${\bf F}_{\rm d}$, on the planet. This force can influence the planet's orbit in two ways: 
(i) Angular momentum: the force creates a torque. In the planar case, the $z$-component of this torque, $\Gamma = {\bf r} \times {\bf F}_{\rm d} \big|_z$, causes the planet's angular momentum to change over time. It is clear that the torque is exerted only by the force component that is perpendicular to the vector ${\bf r}$. In the most general case, the torque is a three-dimensional vector that can lead to changes in the orientation of the orbit or precession, changes in the inclination, and changes in the semi-major axis and eccentricity.
(ii) Energy: the gravitational force transfers energy to or from the planet, causing the planet's orbital energy to change over time, with an energy transfer rate or power $\mathcal{P} = {\bf F}_{\rm d} \cdot {\bf v}_{\rm p}$. Positive power indicates the planet is gaining energy and spiraling outwards, while negative power implies energy loss and the planet spirals inwards.
In mathematical terms, these two effects can be summarized as
\begin{align}
\label{eq:dadt}
\frac{da}{dt} &= -\frac{\mathcal{P}}{E_{\rm p}} a\,, \\
\label{eq:dedt}
\frac{d}{dt}(1-e^2) &= \left(2\frac{\Gamma}{j_{\rm p}} + \frac{\mathcal{P}}{E_{\rm p}}\right) \left(1-e^2\right)\,.
\end{align}
For a planet on a circular orbit, the torque and power are not independent. In this case, the torque is exclusively responsible for changes in the semi-major axis of the planet
\begin{equation}
\frac{da}{dt}= 2\frac{\Gamma}{j_{\rm p}} a\,.
\label{eq:dadt2}
\end{equation}
Eq.\,\eqref{eq:dadt2} shows that whenever a positive or negative torque is applied to the planet, the semi-major axis has to increase or decrease, respectively. 
It is illustrative to consider the case $\Gamma = \xi_0 j_{\rm p}$, with $\xi_0$ a constant, for which the solution of Eq.\,\eqref{eq:dadt2} is an exponential function. Here, the planet's orbit expands or shrinks exponentially due to the external torque on a characteristic time $(2\xi_0)^{-1}$. 
In general, the torque is not strictly proportional to the planet's angular momentum $j_{\rm p}$ (see for example Section \ref{sec:migration-low-mass-planets}). Even if migration is not exponential, it is useful to characterize it through the characteristic migration timescale, $\tau_a$, defined as
\begin{align}
\tau_{a} = \frac{a}{|da/dt|}\,.
\label{eq:migration-timescale}
\end{align}
Equivalent definitions can be derived for the damping/excitation timescales associated with eccentricity and inclination, denoted by $\tau_e$ and $\tau_i$, respectively.

To make progress on the characterization of planet-disk interaction and planet migration, additional information about $\Gamma$ is required. The next sections present how this information can be obtained through either simplified linear or more complex non-linear numerical calculations, and summarize some of the most relevant aspects of the problem.

\subsection{Basic model}

\begin{figure}
\centering
\includegraphics[width=\linewidth]{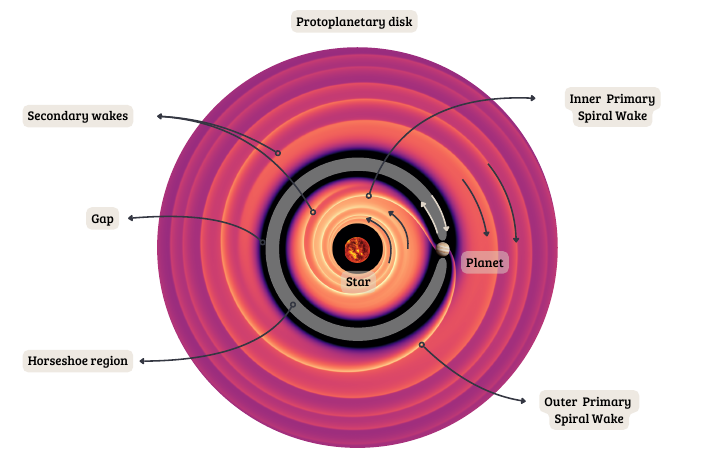}
\caption{Gas surface density obtained after solving the vertically integrated version of equations \eqref{eq:continuity}-\eqref{eq:momentum} numerically considering a massive planet ($m_{\rm p}=10^{-3} M_{\star}$) that is embedded in a thin two-dimensional protoplanetary disk ($h=0.05$).
	The disk temperature is assumed to vary inversely proportional to the radial distance. The planet has a direct (counterclockwise) rotation. Black arrows in the inner and outer regions of the disk represent the sheared velocity field in a frame that corotates with the planet. 
	The brighter the color, the higher the gas density.
	Several features of the planet-disk interaction are visible. 
	(i) High-density primary one-armed spiral wakes launched by the planet that propagate both inward and outward in the disk. These wakes are excited by the planet's gravitational potential and correspond to the natural hydrodynamical response of the gaseous disk to the torque exerted by the planet (see Section \ref{sec:spiral-wakes}). 
	The inner spiral wake, a region of increased density, precedes the planet and exerts a positive torque, propelling it forward in its orbit. Conversely, the outer spiral wake trails the planet and removes angular momentum from it. While secondary, one-armed inner and outer wakes are also present, their launching point is displaced radially and azimuthally relative to the planet compared to the primary wakes.
	(ii) A surface density depletion, or gap, is located within the planetary orbital radius (darkest region). It results from the competition between the planet’s gravitational torque, which pushes gas away from its orbit, and the disk viscous torque, which tends to replenish the gap. In this case, the gap overlaps with a special family of orbits known as
	(iii) horseshoe orbits, depicted by a gray filled region (see Section \ref{sec:horseshoe}). The gas motion in the horseshoe orbits is represented by the gray arrows. Within this region, gas that moves from an outer orbit to an inner one in front of the planet loses angular momentum, which is gained by the planet through a gravitational torque. Conversely, the material behind the planet gains angular momentum, which is lost by the planet. This process is the driver of the non-linear horseshoe drag (see Section \ref{sec:corotation-torques}).
}
\label{fig:planet-disk}
\end{figure}

Given that $\tau_e \ll \tau_a$ for a wide range of physical conditions, most studies of planet-disk interaction focus on determining $\Gamma$. Nevertheless, this simplifying assumption is not strictly needed as both $\Gamma$ and $\mathcal{P}$, or $\tau_a$ and $\tau_e$, can be measured self-consistently from numerical simulations or determined analytically under certain simplifying conditions.
The study of planet-disk interaction and planet migration requires solving the hydrodynamics equations (also known as the Navier-Stokes equations) coupled to the equation of motion of the planet. In addition, non-ideal magneto-hydrodynamics, and radiation hydrodynamics, together with solid dynamics, can also affect the disk response and the resulting force ${\bf F}_{\rm d}$.
If, for simplicity, the effects of solids, magnetic fields, and radiative transfer are neglected, the full set of equations describing the disk dynamics are
\begin{align}
\label{eq:continuity}
\partial_t \rho + \nabla\cdot\left(\rho {\bf v}\right) &= 0\,, \\
\label{eq:momentum}
\partial_t \left(\rho {\bf v} \right) + \nabla\cdot\left(\rho {\bf v}{\bf v} + P{\bf I}\right) &= -\rho \nabla \phi - \nabla\cdot T\,,\\
\label{eq:energy}
\partial_t \epsilon + \nabla\cdot\left(\epsilon {\bf v} \right) &= -P\nabla\cdot {\bf v}\,,
\end{align}
where $\rho$, ${\bf v}$, $P$, and $\epsilon$ are the gas density, velocity, pressure, and internal energy density, respectively. While molecular viscosity is negligible in protoplanetary disks, the viscous stress tensor, $T$, is typically included as a sub-grid model for turbulent viscosity.
In an astrocentric frame of reference, the gravitational potential, $\phi$, is given by
\begin{equation}
\label{eq:potential}
\phi = -\underbrace{\frac{GM_\star}{r}}_{\text{\normalfont Star}}  - \underbrace{\frac{Gm_{\rm p} }{|{\bf r}-{\bf r}_{\rm p}|}}_{\text{\normalfont Planet}} + \underbrace{\frac{Gm_{\rm p} \, ({\bf r} \cdot {\bf r}_{\rm p})}{r_{\rm p}^3}}_{\text{\normalfont Indirect potential}}\,,
\end{equation}
where, for simplicity, only one star at the center of the frame of reference and one planet at ${\bf r}_{\rm p}$ are considered. The first and second terms correspond to the central potential of the star and planet, respectively. The third one, the indirect potential, is a non-inertial term that arise from the choice of an astrocentric frame of reference.

Self-consistent migration studies also require solving the equation(s) of motion of the planet(s), which is(are) coupled to the dynamical evolution of the protoplanetary disk
\begin{equation}
\label{eq:planet}
\frac{d^2 {\bf r}_{\rm p}}{dt^2} = -\frac{GM_\star}{r_{\rm p}^2} \hat{r} + \frac{{\bf F}_{\rm d}}{m_{\rm p}}\,.
\end{equation}
The gravitational force exerted by the disk, ${\bf F}_{\rm d}$, depends on the density distribution $\rho$, and it is given by
\begin{equation}
\label{eq:force_disk}
{\bf F}_{\rm d} = G m_{\rm p} \int_{\rm disk} \rho \frac{{\bf r}-{\bf r}_{\rm p}}{{|{\bf r}-{\bf r}_{\rm p}|}^3} dV\,,
\end{equation}
with $dV$ the volume element. Here, the fluid indirect acceleration and self-gravity are neglected. These effects can be relevant when protoplanetary disks are massive enough (for example, disks around young systems), in specific regions with high gas/solids density, or around planets with massive circumplanetary disks.

To close the system, a relationship between the gas pressure, the gas density, and internal energy density is needed. 
Assuming an ideal equation of state for the gas, the pressure and the energy density are related as
\begin{equation}
\label{eq:adiabatic}
P = \left(\gamma - 1\right) \epsilon\,,
\end{equation}
where $\gamma$ is the adiabatic index, equal to $5/3$ for atomic gas. If the gas is locally isothermal, the energy equation \eqref{eq:energy} disappears from the problem, and the pressure is given by
\begin{equation}
\label{eq:isothermal}
P = c_{\rm s}^2 \rho \,,
\end{equation}
with $c_{\rm s}$ the sound speed. Eq.\,\eqref{eq:isothermal} is a particular form of a barotropic equation of state, $P = P(\rho)$. 

If the protoplanetary disk is thin, azimuthally symmetric (or axisymmetric), vertically isothermal (i.e., its temperature does not depend on altitude), and it is in vertical hydrostatic equilibrium (i.e., the vertical gravity of the central star is balanced by the vertical pressure gradient), the gas density can be approximated by
\begin{equation}
\rho(r,z) \simeq \rho_0(r) e^{-z^2/(2H)^2}\,,
\end{equation}
where the scale-height of the disk is $H = c_{\rm s}/\Omega_{\rm K}$, with $\Omega_{\rm K} = \sqrt{GM_\star/r^3}$ the Keplerian angular velocity.
Given that in this simple model the planet's potential is neglected, the scale-height depends only on the radial distance.
Also, because the viscosity in protoplanetary disks is expected to be low, the disk radial velocity is small (at least in the disk midplane).
The balance between the stellar gravitational gradient, the gas pressure gradient, and the centrifugal force produces a disk that rotates at a sub-Keplerian speed.
This disk model is a reasonable one to use as an initial condition or to describe the global disk structure as long as the planet's potential does not significantly affect the disk structure.
This is not the case for massive planets (see e.g., Section \ref{sec:gap}). 
Two disk models are widely used when the temperature is prescribed externally: one in which the disk is vertically isothermal on curves of constant cylindrical radius \citep[see e.g.,][]{2013MNRAS.435.2610N} and one in which the disk is vertically isothermal on curves of constant spherical radius \citep[see e.g.,][]{2016ApJ...817...19M}.

The viscosity model that is mostly used in planet-disk interaction studies is the so-called $\alpha$-viscosity model.
This viscosity prescription assumes a kinematic viscosity $\nu = \alpha_\nu c_{\rm s} H$, with $\alpha_\nu \ll 1$ a dimensionless constant ($\alpha_\nu \sim 10^{-5}-10^{-3}$). 
The simplicity of this model makes it a practical choice for planet-disk interaction studies, even if it may not always achieve the highest levels of precision or realism.

Protoplanetary disks are relatively cold thin disks, with an aspect-ratio $h \equiv H/r\ll 1$. Typical values for $h$ are within the range $h\sim 0.03-0.07$, and depending on the thermal structure of the disk it can vary both locally and globally. Because protoplanetary disks are thin disks, a widely used simplification is to assume them as two-dimensional structures and neglect their vertical motion. 
In this case, the three-dimensional Navier-Stokes equations are replaced by their vertically integrated version, for which the density is replaced by the surface density, $\Sigma=\int_{-\infty}^{\infty}\rho\, dz$, and the planet's potential is replaced by the Plummer potential
\begin{equation}
\phi_{\rm p} = -\frac{Gm_{\rm p}}{\sqrt{|{\bf r}-{\bf r}_{\rm p}|^2+r_{\rm s}^2}}\,,
\end{equation}
with $r_{\rm s}$ a smoothing or softening length, typically proportional to the local scale height of the disk ($r_{\rm s} \sim 0.6 H(r_{\rm p})$). The effect of the smoothing length is to reduce the effective potential close to the planet and make it feel a disk torque closer to what it would feel from a three-dimensional disk.
In this two-dimensional approximation, the disk's force is calculated with Eq.\,\eqref{eq:force_disk}, after replacing the density, $\rho$, by the surface density, $\Sigma$, and the volume element, $dV$, by the area element, $dA$.

Solving Eqs.\,\eqref{eq:continuity}-\eqref{eq:energy} coupled to Eq.\,\eqref{eq:planet} and Eq.\,\eqref{eq:adiabatic} or Eq.\,\eqref{eq:isothermal} from an initial condition with appropriate boundary conditions corresponds to a self-consistent planet-disk interaction and planet migration study. 
Finding the solution of the system involves obtaining the planet distance and velocity as a function of time together with the disk structure and dynamics (i.e., the disk density, velocity vector, and temperature as a function of space and time).

\subsubsection{Disk evolution}

Planets migrate within protoplanetary disks, which are dynamic structures that evolve. These disks are relatively short-lived, typically dissipating within 1-10 million years due to various processes. Viscous accretion, magneto-hydrodynamical winds, and photoevaporation are among the primary mechanisms driving disk evolution. While the first two transport angular momentum within the disk, photoevaporation is unique in that it disperses the disk without affecting its angular momentum distribution.
Photoevaporation is a form of a thermal outflow driven by the radiation absorbed by the disk gas from the central star (internal photoevaporation) or a nearby companion (external photoevaporation). The disk's surface layers, directly exposed to this radiation, become heated to the point where the gas can escape the system. 
Interestingly, photoevaporation models predict a mass loss profile that is maximum at some specific radius in the disk. If viscous evolution (or any other mechanism) is not able to provide enough mass to compensate for the mass lost by evaporation, a gap would form initially close to this radius. 
Gap formation is a common feature of photoevaporation, and these gaps can have important consequences not only for planet migration, as they create large vortensity gradients and strong corotation torques (see Section \ref{sec:corotation-torques}), but also for the distribution of solids within the disk, which are the building blocks of planets.
In the framework of viscous disk model, the simplest equation describing the evolution of an axisymmetric vertically integrated Keplerian disk arises from mass and momentum conservation laws. In this case, the disk evolution can be modeled as
\begin{equation}
\frac{\partial \Sigma}{\partial t} - \frac{3}{r}\frac{\partial}{\partial r} \left[\sqrt{r} \frac{\partial}{\partial r} \left( \sqrt{r} \nu \Sigma \right) \right] = \dot{\Sigma}_{\rm w}(r,t)\,,
\end{equation}
where the second term in the L.H.S. corresponds to mass transport by viscous stress and $\dot{\Sigma}_{\rm w}$ models the mass loss due to winds. Additional terms accounting for angular momentum/mass transport can also be included in this simplified model. These terms depend on $r,t$ and have to be modeled carefully through detailed numerical simulations, including detailed thermodynamics, magneto-hydrodynamics, and chemistry.
The way the disk dissipates, and when this happens, plays a crucial role in determining how planets could form and what would be the final architecture of planetary systems. Therefore, a unified picture of planet formation must involve a precise understanding of the physics and timing of the different processes affecting the disk evolution. 
Once the disk dissipates, the so-called planet-disk interaction stops. 

Further reading about the disk evolution by the different mechanisms mentioned here can be found in the recent review article by \cite{2023ASPC..534..567P}, and references therein. 

\subsection{Main features of planet-disk interaction}
\label{sec:main-features}

The main features of the planet-disk interaction can be presented by considering the simple case of a single planet on a circular direct orbit embedded in a vertically integrated gaseous disk that rotates differentially at a velocity proportional to the Keplerian velocity. 
A numerical solution of this problem is shown in Fig.\,\ref{fig:planet-disk}. This figure shows a snapshot of the surface density obtained after solving the Navier-Stokes equations \eqref{eq:continuity}-\eqref{eq:momentum} numerically for a viscous ($\alpha_\nu=3\times 10^{-3}$) locally isothermal disk ($h=0.05$) with a Jupiter-like embedded planet ($m_{\rm p} = 10^{-3} M_\star$).
While only surface density perturbations are shown, the azimuthal and radial velocities (as well as energy density or disk temperature in a non-isothermal case), are also non-trivial fields that must be solved self-consistently.
The gravitational force exerted by the disk on the planet (see Eq.\ref{eq:force_disk}) arises from the disk response to the gravitational potential of the planet.
The following features in the gas can exert gravitational forces onto the planet and modify its orbit: (i) inner and outer spiral wakes that can be primary, secondary, etc., (Section \ref{sec:spiral-wakes}) (ii) horseshoe orbits (Section \ref{sec:horseshoe}, \ref{sec:corotation-torques}, \ref{sec:solids}) and (iii) Gas depletion or gap at the planetary orbit (Sections \ref{sec:gap} and \ref{sec:migration-massive-planets}). In addition, small- and large-scale vortices (not present in Fig.\,\ref{fig:planet-disk}) are prone to develop if planets are massive enough and/or disk viscosity is low enough (see Section \ref{sec:vortices}).

\subsubsection{Spiral wakes}

\label{sec:spiral-wakes}
\begin{figure}
\centering
\includegraphics[width=1\linewidth]{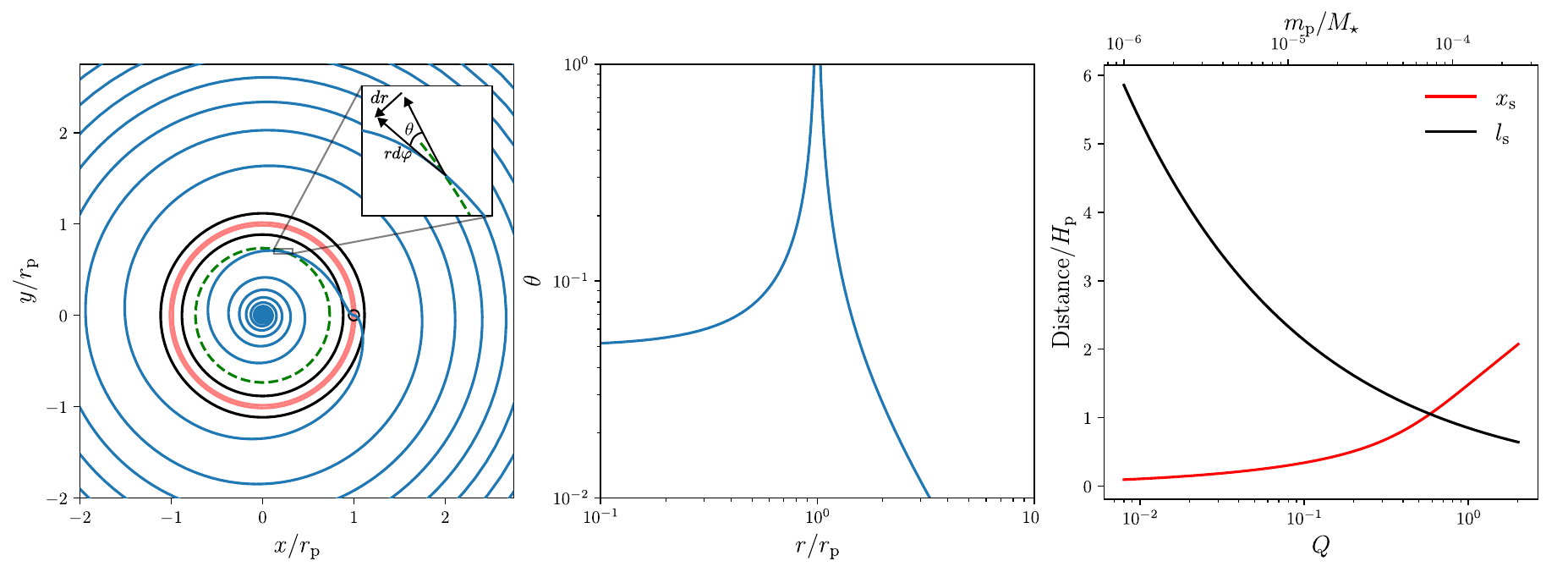}
\caption{Main geometrical features of the spiral density wake, horseshoe region and shock distance (see Section \ref{sec:spiral-wakes}, \ref{sec:horseshoe}). The left panel shows the inner and outer spiral wake calculated from Eq.\,\eqref{eq:wake} for a disk with $H_{\rm p}=0.05 r_{\rm p}$. Green dashed line corresponds to a reference circle to measure the spiral wake pitch angle $\theta$ (see zoom-in). The red solid lines correspond to the horse-shoe semi-width to each side of the planet (see Eq.\,\ref{eq:horseshoe-xs}), and the black solid lines are the shock distance (see Eq.\,\ref{eq:lshock}). Calculations were done for a planet of mass $m_{\rm p} = 10^{-5} M_\star$.
	The middle panel shows the pitch angle (Eq.\,\ref{eq:wake-pitch}) with radius. Close to the planet, as the wake emerges radially, the pitch angle diverges. For large radii, $\theta \to 0$, while in the inner regions of the disk, $\theta \to h$.
	Right panel shows the dependence of the horseshoe semi-width, $x_{\rm s}$, and the shock distance, $l_{\rm s}$, with $Q$, the planet mass relative to the thermal mass.}
\label{fig:wake}
\end{figure}

Right after the planet is inserted in the disk, one-armed inner and outer spiral wakes emerge from the planet and corotate with it. Therefore they are stationary structures in a corotating frame.
In two dimensions, the wake is interpreted as the constructive interference of the waves launched at the Lindblad resonances (see Section \ref{sec:migration-low-mass-planets}) by each Fourier component of the planet potential \citep{2002MNRAS.330..950O}. For a low-mass planet (where low means low enough so linear theory is valid, see below and Section \ref{sec:migration-low-mass-planets} for its definition), the peak density of the wake is approximately given by
\begin{equation}
\varphi_{\rm wake}(r) \simeq \varphi_{\rm p} + {\rm sign}\left(r-r_{\rm p}\right)\int_{r_{\rm p}}^{r} \frac{\Omega - \Omega_{\rm p} }{c_{\rm s}} dr'\,,
\label{eq:wake}
\end{equation}
where $\Omega$ and $\Omega_{\rm p}$ are disk and planet angular velocities, respectively. Eq.\,\eqref{eq:wake} shows that the wake's peak-density propagation is consistent with an acoustic perturbation originating from the planet that propagates away at the local sound speed while being transported azimuthally by the shear (Keplerian) flow. Therefore, the relevant timescale for the wake propagation is the sound-crossing time, which is quite short.
The leftmost panel of Fig.\,\ref{fig:wake} depicts the inner and outer spiral wakes (blue curves), as predicted by Eq.\,\eqref{eq:wake} for a locally-isothermal disk with aspect-ratio h=0.05. Initially, the wake propagates radially from the planet (open circle) but subsequently becomes twisted as it moves away due to Keplerian shear. This can be quantified precisely through the wake pitch angle, $\theta$.
The wake pitch angle, defined as the angle between the spiral pattern and a circle with the same center, can be obtained from Eq.\,\eqref{eq:wake}
\begin{align}
\tan(\theta) = \frac{1}{r}\frac{dr}{d\varphi} = \frac{1}{r} \frac{c_s}{\Omega-\Omega_{\rm p}}\,.
\label{eq:wake-pitch}
\end{align}
In the leftmost panel of Fig.\,\ref{fig:wake} the pitch angle is represented geometrically by the angle formed between the blue and green dashed curves at their intersection (see zoom-in).
In addition, the middle panel of Fig.\,\ref{fig:wake} shows the wake pitch angle as a function of distance. 
In the outer reaches of the disk, the wake pitch angle becomes negligible. There, the wake is a tightly wound spiral, with a winding level that increases with radius. The wake winding rate depends on the sound speed, where disks characterized by higher temperatures produce less tightly wound wakes. In the inner disk, the pitch angle reaches the fixed value $h$.

Far enough from the planet and in the absence of dissipation (e.g., viscosity), the spiral waves evolve to the point in which non-linear effects become relevant and they experience what is known as profile steepening, leading to a shock formation \citep{2001ApJ...552..793G}, producing a discontinuity in the flow.
The distance from the planet at which the shock develops, $l_{\rm s}$, is approximately given by
\begin{equation}
l_{\rm s} \simeq 1.12 H_{\rm p} \left(\gamma+1\right)^{-2/5} Q^{-2/5}\,,
\label{eq:lshock}
\end{equation}
where $Q = m_{\rm p}/(M_\star h^3)$ is the ratio between the planet mass and the so-called thermal mass, $m_{\rm th} = h^3 M_{\rm \star}$.
The left panel of Fig.\,\ref{fig:wake} shows with solid black lines the location of shock formation for a planet of mass $m_{\rm p} = 10^{-5} M_\star$. As long as $m_{\rm p} \lesssim m_{\rm th}$, the shock distance, $l_s$, is larger than the pressure scale of the disk, $H_{\rm p}$, and linear waves can be exited.
In addition, in this regime, the shock distance is larger than the semi-width of the horseshoes depicted with red lines (see Section \ref{sec:horseshoe}). For the planet mass and disk aspect ratio considered, the shock develops outside of the horseshoe region.

The right panel of Fig.\,\ref{fig:wake} shows the shock distance (black solid line) and the horseshoe semi-width (red solid line) as a function of $Q$.
The mass for which both distances are equal is typically the limiting mass for which one can assume the planet to be low-mass and in a linear regime.

Still, wake-like perturbations are more complex. Far enough from the planet, the interference of additional azimuthal modes causes the appearance of additional spiral wakes, known as secondary, tertiary, etc. For low-mass planets, the inner wake is a collection of several narrowly spaced spiral arms \citep{2018ApJ...859..118B}. 
These additional wakes are not seen in Fig.\,\ref{fig:planet-disk} because it shows the case of a massive planet that carves a gap (see Section\,\ref{sec:migration-massive-planets}), but can be recognized in the innermost part of the disk for left and middle panels of Fig.\,\ref{fig:horseshoe}. As the planet's mass increases, the secondary wake becomes more prominent, and its separation from the primary increases \citep{2015ApJ...815L..21F}. This secondary wake that is significantly offset in azimuth to the primary one is visible in Fig.\,\ref{fig:planet-disk}. This effect is also recognized after a visual inspection of the inner disk in each panel of Fig.\,\ref{fig:horseshoe}.

\subsubsection{Horseshoe region}
\label{sec:horseshoe}

\begin{figure}
\centering
\includegraphics[width=1.0\linewidth]{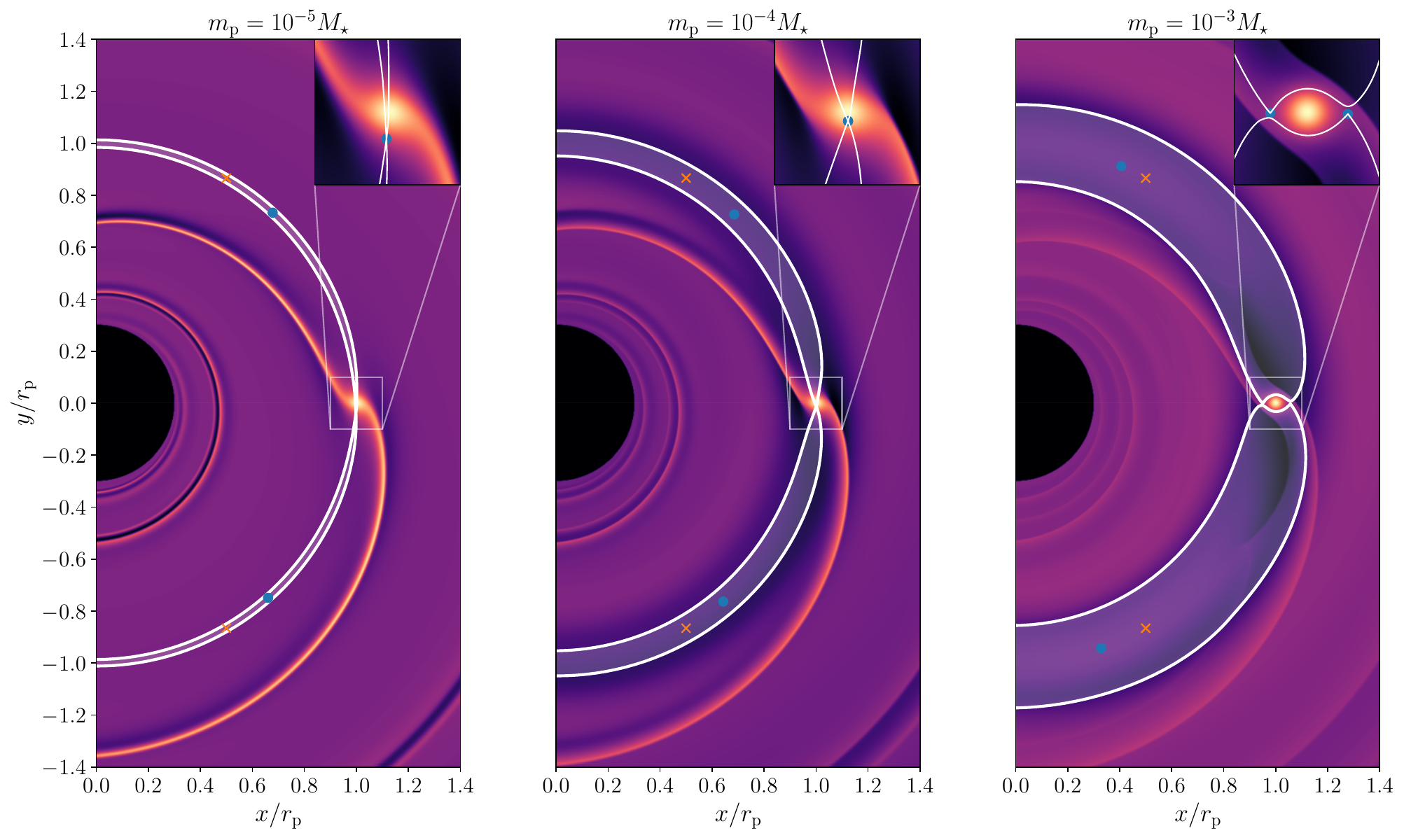}
\caption{Gas surface density obtained for planets of mass $10^{-5} M_{\star}$ (left panel), $10^{-4} M_{\star}$ (middle panel) and $10^{-3} M_{\star}$ (right panel). The planet is located at $(x/r_{\rm p},y/r_{\rm p}) = (1,0)$. The disk aspect-ratio is $h=0.05$ and viscosity is large ($\alpha_\nu = 10^{-3}$). 
	For each case, the horseshoe region is depicted with a shaded region delimited by white thick lines (horseshoe separatrix). The stagnation points calculated for these particular snapshots are depicted with blue circles. Outside of these lines, the material keeps orbiting or circulating the star while being slightly perturbed by the planet. Material within the horseshoe region librates, performing U-turns when approaching the planet. 
	Inspection of the zoom-in panels shows that for low-mass planets (left and middle panels), a single stagnation point exists where the thick white lines intersect. For larger planets, the region around the planet expands, allowing the existence of additional stagnation points, like L1 and L2 for the restricted three-body problem.
	In all the cases, it is also possible to have stagnation points analogous to the triangular L4 and L5 Lagrange points of the restricted three-body problem, that are drawn with orange crosses as a reference. A hydrodynamical version of the L3 Lagrange point also is present in all these cases but it is out of the plotted region.
	Another feature observed is that the width of the horseshoe increases with $m_{\rm p}$. Also, the topology of horseshoes changes as $m_{\rm p}$ increases. For sufficiently large masses (rightmost panel), the single stagnation point transforms into two points that move away from the planet as its mass increases.  
}
\label{fig:horseshoe}
\end{figure}

In the framework of the restricted three-body problem, test particles that have small eccentricity and a semi-major axis that is close to the planet describe a motion that is similar to a horseshoe. These are the so-called horseshoe orbits \citep[see e.g.,][]{1999ssd..book.....M}. 
When test particles are replaced by gas, horseshoe orbits are still possible, but they are, in general, different for the same primary and secondary masses. 
This is because not only the gravitational potential plays a significant role, but also the pressure gradient (see Eq.\,\ref{eq:momentum}).
The topology of horseshoe orbits is defined by the number and location of the stagnation points. A stagnation point is a point where the velocity of the fluid, in a corotating frame with the planet, is zero.
In the restricted three-body problem, there are five stagnation points known as the Lagrange equilibrium points. It is worth mentioning that the indirect potential of Eq.\,\eqref{eq:potential} is crucial for the existence of the so-called triangular points, both in the restricted three-body problem as well as in its hydrodynamical version.
In the gaseous case, depending on the mass of the perturber, both the location and the number of stagnation points can differ from the restricted three-body problem. In general, its number depends on the mass of the secondary, and its position depends on the disk physics and response to the planet perturbation. For planets that are massive enough, gravity dominates over the pressure, and stagnation points that are analogous to the L1 and L2 Lagrange points can exist. In addition, L3, L4, and L5 are present whenever the indirect potential is not neglected.
Sufficiently close to a low-mass planet there is only one stagnation point that, depending on the local pressure gradient, can be behind or ahead of the planet azimuthally \citep[see e.g.,][]{2006ApJ...652..730M}, which can affect significantly the net torque felt by the planet.

In Fig.\,\ref{fig:planet-disk}, the horseshoe region is schematically depicted with a gray-shaded region. The motion of gas along the horseshoes is represented by two gray arrows. 
Within this region, gas moving from an outer orbit to an inner one in front of the planet loses angular momentum, which, by angular momentum conservation, must be gained by the planet through a gravitational torque. Conversely, the material behind the planet gains angular momentum, which is lost by the planet. 
If the mass is distributed evenly within the horseshoe and the orbits are fully symmetrical like the ones in the restricted three-body problem, the net torque averages to zero over a full horseshoe cycle or libration timescale. However, hydrodynamical effects make the horseshoe non-symmetric, enabling the existence of a net torque arising from the dynamics. As will be seen later, the quantity that controls the existence of torque from within the horseshoe region is the disk vorticity gradient divided by the disk surface density.

The horseshoe libration timescale corresponds to the synodic period of a particle located at a distance equal to the horseshoe semi-width $x_{\rm s}$. In a frame corotating with the planet and sufficiently close to $r=r_{\rm p}$, the Keplerian speed can be expanded in series to first order in $\Delta r=r-r_{\rm p}$ as $-3/2 \Omega_{\rm p} \Delta r$.
Therefore, the synodic period of a particle in the horseshoe is approximately equal to $4\pi r_{\rm p}/(3 \Omega_{\rm p} x_s)$. However, the time at which a fluid particle would be at the same location with respect to the planet is twice the synodic period because of the U-turn motion in front or behind the planet. Therefore, the horseshoe libration timescale is
\begin{equation}
\tau_{\rm libration} = \frac{8\pi r_{\rm p}}{3\Omega_{\rm p} x_{\rm s}}\,.
\label{eq:horseshoe-libration-timescale}
\end{equation}
This is an important timescale for planet-disk interaction as it informs the timescale at which material that is close to the planet can get redistributed within the horseshoe region. Processes that affect the horseshoe material on shorter timescales can disrupt the horseshoe motion (see e.g., Section \ref{sec:solids}).

The radial extent of the outermost horseshoe orbit (separatrix) or horseshoe semi-width, $x_{\rm s}$, depends on the planet's mass, its distance to the central star, and the disk temperature. The half-width of the horseshoe region, $x_{\rm s}$, valid for a wide range of masses, is given by \cite{2017MNRAS.471.4917J}
\begin{equation}
x_{\rm s} = H_{\rm p} \frac{1.05 Q^{1/2} + 3.4 Q^{7/3}}{1+2Q^2} = H_{\rm p}\begin{cases}
	1.05 Q^{1/2} \quad & Q\ll 1 \\
	1.7 Q^{1/3} \quad & Q\gg 1 
\end{cases}\,.
\label{eq:horseshoe-xs}
\end{equation}
In the limit of large planetary masses (where large is relative to the thermal mass $m_{\rm th}$), the scaling follows that of the restricted three-body problem ($m_{\rm p}^{1/3}$). For low-mass planets, $x_{\rm s}$ scales with the square root of the planetary mass.
The same scaling but different coefficients are found in two-dimensional simulations, and the width depends on the smoothing length, $r_{\rm s}$, adopted \citep{2009MNRAS.394.2297P}. Globally isothermal three-dimensional simulations of low-mass planets show negligible width dependence on height -\citep[see e.g.,][]{2016ApJ...817...19M}, while for the case of adiabatic disks, the horseshoes gradually shrink with height \citep{2019ApJ...887..152F}.

Fig.\,\ref{fig:horseshoe} shows the gas surface density obtained numerically for different planet masses embedded in a thin viscous disk. For all planet masses, white thick solid lines delineate the horseshoe separatrix. In all the cases, stagnation points are calculated and depicted with blue circles. Zoom-in panels in the same figure show that, depending on the planet mass, one or two stagnation points can exist close to the planet. Also, the figure shows the hydrodynamical version of the L4 and L5 Lagrange points that are azimuthally shifted to L4 and L5 in the restricted three-body problem. After inspecting all the panels, it is also clear a dependence of the horseshoe width with the planet mass (Eq.\,\ref{eq:horseshoe-xs}).

After using Eq.\,\eqref{eq:lshock} and \eqref{eq:horseshoe-xs}, the value of $Q$ for which the horseshoe semi-width, $x_{\rm s}$, matches the shock formation distance, $l_{\rm s}$, is $Q\simeq 0.6$, and it occurs at a distance $\simeq H_{\rm p}$ (see the third panel of Fig.\,\ref{fig:wake}). For larger values of $Q$, $l_{\rm s}<x_{\rm s}$ and non-linear effects in the planet vicinity can be relevant.

\subsubsection{Planet gap}
\label{sec:gap}

Planets that are massive enough to repel part of the disk mass from their horseshoe region create a density depletion or gap in the disk which modifies substantially how they migrate (see Section \ref{sec:migration-massive-planets}).
Here, massive means a planet with sufficient mass to create a distinct gap in the disk but not so much as to dramatically change the disk overall dynamics within a relatively short period.
For typical disk parameters, a $\sim$ Jupiter-like planet could be considered as a limiting case. For example, for planets larger than a few $10^{-3} M_\star$ the disk can become eccentric (the precise value depends on disk viscosity and or aspect-ratio).
Not only the mass but the disk viscosity and disk temperature (or disk scale height) play a significant role in the onset of the gap and the corresponding disk response to the planet perturbation.
Numerical simulations show that the azimuthally-averaged surface density at the planet radius is well approximated by $\Sigma_{\rm gap} \simeq {\Sigma_0} \left(1+ s K\right)^{-1}$, with $s\simeq0.04$ a small constant, $\Sigma_0$ the initial unperturbed density and $\displaystyle{K = \left(m_{\rm p}/M_\star\right)^2 h^{-5} \alpha_{\nu}^{-1}}$ \citep[see e.g.,][]{2017PASJ...69...97K}.
Therefore, the planetary mass needed to carve a gap whose density at the bottom is a fraction, $f_{\rm gap}$, of the unperturbed initial surface density is
\begin{equation}
m_{\rm p}\simeq s^{-1/2} \left(\frac{1-f_{\rm gap}}{f_{\rm gap}}\right)^{1/2} h^{5/2} \alpha_{\nu}^{1/2} M_\star\,.
\end{equation}
This gap results from the balance between the torque deposited in the disk by the planet through shocks at specific locations, which makes the gas move away from the planet, and the disk viscous torque, which acts to flatten the gap created by the planet. In particular, low-mass planets embedded in an inviscid disk can deposit angular momentum at the shock distance $l_{\rm s}$ (see Eq.\,\ref{eq:lshock}) and create a gap (as long as they keep in the same radial distance for a sufficiently long time).
Given that the gap and resulting disk structure evolve on viscous time scales ($\tau_{\rm \nu} \sim r^2/\nu$), boundary conditions would affect the resulting disk/gap structure \citep[see e.g.,][]{2020ApJ...891..108D}. Therefore, the development of realistic boundary conditions is key to studying this important feature of planet-disk interaction and the resulting disk and planet dynamics. 
Further progress on gap modeling would require incorporating the gap dependence with the migration rate, planet accretion, and turbulence together with realistic boundary conditions.

\subsubsection{Vortices}
\label{sec:vortices}

\begin{figure}
\centering
\includegraphics[width=\linewidth]{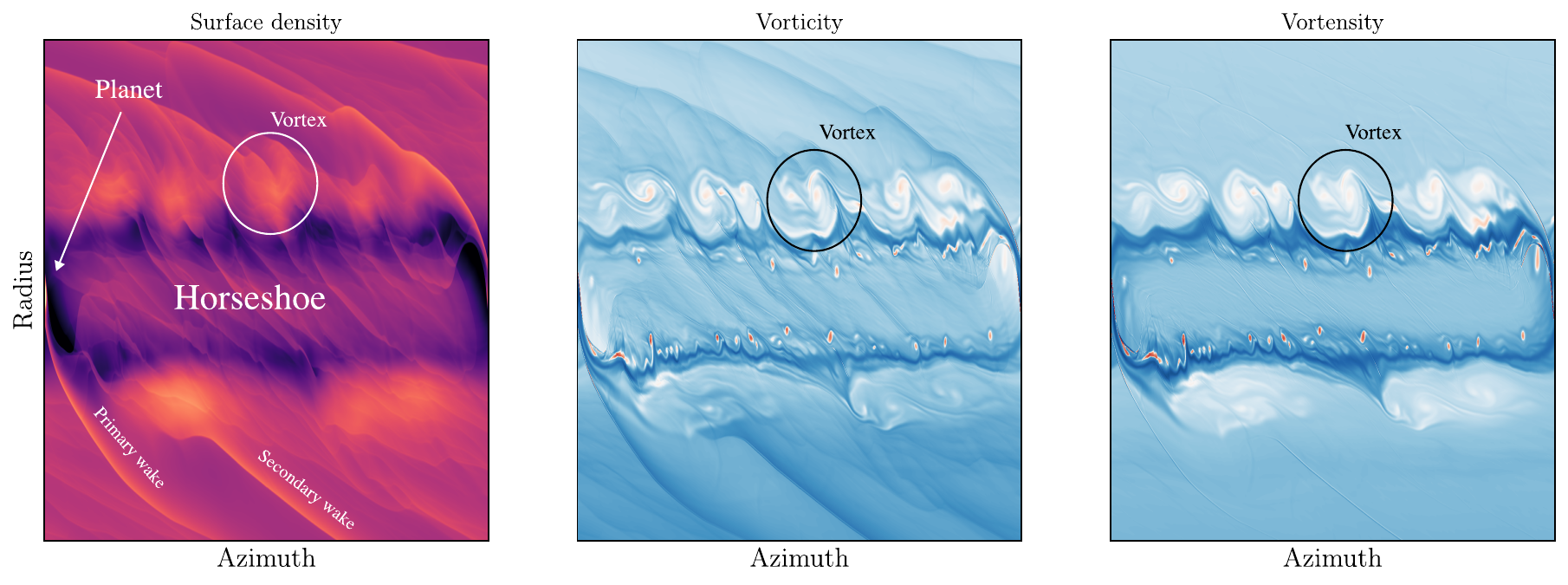}
\caption{Surface density, vorticity, and vortensity in polar coordinates from a numerical simulation of a planet embedded in a low-viscosity disk of aspect-ratio $h=0.035$. The planet mass is $m_{\rm p} = 5\times 10^{-4} M_\star$. 
	The snapshot was taken at 20 planetary orbits after inserting the planet. This time scale corresponds to an early phase in which a first generation of vortices is created.
	Vertical axes are in logarithmic scale from $r/r_{\rm p}\in[0.61,1.67]$. The surface density is plotted relative to the initial density profile in logarithmic scale, in the range $[10^{-1},10]$. The vorticity and vortensity are displayed in a logarithmic scale in the range $[-10,10]$, $[-20,20]$ with a small linear transition from negative to positive values.
	In addition, within the horseshoe region, a gap is developing (see Section\,\ref{sec:gap}).
}
\label{fig:vortices}
\end{figure}

When the disk viscosity is low or the planetary mass is large enough, vortices in the gas can be triggered by the presence of the planet. A vortex is a region of the fluid that spins rapidly around a central point (the vortex center), creating a low-pressure region.
One crucial mechanism to excite vortices is known as the Rossby wave instability \citep{1999ApJ...513..805L}.
This linear instability is typically triggered by localized $\sim$ azimuthally symmetric bumps in gas density (or pressure). The linear phase is followed by the non-linear evolution of the instability for which smaller vortices can merge and form larger vortices. In the context of planet-disk interaction, this process typically occurs near the edge of a gap carved by a planet, where the pressure has a maximum.

A good indicator for the presence of vortices in the disk is the $z$-component of the vorticity $(\nabla \times {\bf v})_z$ or vortensity $(\nabla \times {\bf v})_{\rm z}/\Sigma$ (see also Section \ref{sec:corotation-torques}). As an example of the relevance of these quantities, 
Fig.\,\ref{fig:vortices} shows a snapshot of surface density, vorticity, and vortensity in polar coordinates from a numerical simulation of an intermediate-mass planet ($m_{\rm p}=5\times 10^{-4} M_\star$) embedded in a thin ($h=0.035$) low-viscosity ($\alpha_\nu=10^{-4}$) disk, where the development of vortices triggered by the Rossby wave instability is expected.
It is also clear the development of spiral wakes triggered by vortices. 
While spiral wakes are nearly imperceptible in the vortensity map, prominent features such as vortices (red spots with large vorticity/vortensity) and shocks (see e.g., the narrow stripes near the planet) are readily identifiable. This can be understood from the fact that vortensity is conserved along fluid streamlines in two-dimensional inviscid barotropic fluids. Therefore, it cannot display the passage of acoustic or density waves. As such, a vortensity map is similar to a vorticity map where the effect of waves is filtered out. Furthermore, the comparison of vorticity and vortensity maps allows the waves to be highlighted.
Vortex behavior, including merging, growth, dissipation, and migration varies based on viscosity levels.

\subsection{Migration of low-mass planets}
\label{sec:migration-low-mass-planets}
For planetary masses that are small enough, the response of a protoplanetary disk to the planet's gravitational potential can be studied in the linear regime. Here, small means much smaller than the thermal mass $m_{\rm th}$, for which the development and propagation of linear waves is possible on scales larger than the disk pressure scale, $H$.
Given the azimuthal periodicity of the system, the planetary potential is decomposed azimuthally in Fourier series. For example, in the case of a single planet on a circular orbit in a two-dimensional (or vertically integrated) disk, the potential can be written as \citep{1979ApJ...233..857G}
\begin{equation}
\phi_{\rm p}(r,\varphi,t) = \sum_{m=0}^{\infty} \phi_m(r) \cos\left[m\left(\varphi - \Omega_{\rm p} t\right)\right]\,,
\end{equation}
where only even terms (cosines) are needed due to the parity of the planetary potential. After inserting this potential in Eqs.\,\eqref{eq:continuity}-\eqref{eq:momentum} and applying linear theory, one finds the resulting equations to be singular at some specific radial locations or resonances. All these resonant radii are the places in the disk where strong perturbations and torques are exerted. The location of the resonances is given implicitly by the condition
\begin{equation}
m\left(\Omega-\Omega_{\rm p}\right) = \begin{cases} 
	\pm \kappa &\quad (\rm Lindblad ~ resonances) \\
	0 &\quad (\rm Corotation ~ resonance)\\
\end{cases}
\label{eq:resonances}
\end{equation}
where $\kappa$ is the natural frequency at which a small radial disturbance in the disk oscillates, called epicyclic frequency. 
For nearly-Keplerian protoplanetary disks, $\kappa \simeq \Omega$, therefore the nominal radial location of the Lindblad resonances, $r_{\rm L}$, is
\begin{equation}
r_{L} = \left(\frac{m}{m \pm 1}\right)^{2/3} r_{\rm p}\,.
\label{eq:lindblad-resonance}
\end{equation}
The minus (plus) sign in Eq.\,\eqref{eq:lindblad-resonance} defines the Outer (Inner) Lindblad resonances. 
The corotation resonance corresponds to the radius where the disk is at rest with respect to the planet. For a strictly Keplerian disk and a planet on a circular orbit, this is exactly the planet's orbital radius or semi-major axis.
The location of the Lindblad resonances, given by Eq.\,\eqref{eq:lindblad-resonance}, depends on the disk model and physics. One important result is that when one considers the effect of pressure gradients, for $m\to\infty$ the effective location of the Lindblad resonances tends to $r_{\rm p} \pm 2/3 H$ instead of $r_{\rm p}$ as predicted by Eq.\,\eqref{eq:lindblad-resonance} \citep{1993ApJ...419..155A}, which defines the minimal distance from the planet at which waves can be triggered.

Linear and non-linear numerical calculations demonstrate that the total torque felt by a planet in locally isothermal disks can be written as
\begin{equation}
\Gamma\left(r_{\rm p}\right) = \psi\left(\alpha,\beta\right)\, \Gamma_{\rm p}\,,
\label{eq:total-torque}
\end{equation}
where $\psi = \psi(\alpha, \beta)$ is an $\mathcal{O}(1)$ function that consists of a linear combination of the exponents of the surface density and temperature power laws, given by $\alpha = -d\log \Sigma/d\log r$ and $\beta = -d\log T/d\log r$. 
The factor $\Gamma_{\rm p}$ is a reference torque at the planet position, given by
\begin{equation}
\Gamma_{\rm p} = \left(\frac{m_{\rm p}}{M_{\star}}\right)^2 h_{\rm p}^{-2} \Sigma_{\rm p} r_{\rm p}^4 \Omega_{\rm p}^2\,.
\label{eq:gamma0}
\end{equation}
where all the factors must be evaluated at the planet's orbital radius $r_{\rm p}$. 
The surface density $\Sigma_{\rm p}$ corresponds to the unperturbed disk surface density. Low-mass planets that migrate due to the torque of Eq.\,\eqref{eq:total-torque} are said to experience type-I migration.
The most recent calculation of the total torque exerted by the disk onto a low-mass planet was done by \cite{2024ApJ...968...28T}, showing that, for a three-dimensional locally isothermal disk, the function $\psi$ is
\begin{equation}
\psi(\alpha,\beta) = -\left(1.436 + 0.537\alpha + 0.439\beta\right)\,.
\end{equation}

With this information, it is now possible to address the question raised in Section \ref{sec:planet-disk-migration} concerning the conditions under which an exponential or self-similar migration might occur.
After combining Eqs.\,\eqref{eq:migration-timescale} and \eqref{eq:total-torque}, the migration rate is
\begin{equation}
\tau_{\rm a} = \frac{1}{2|\psi|} \left(\frac{m_{\rm p}}{M_\star}\right)^{-2}  \left(\frac{h_{\rm p}^2 \Omega_{\rm p}^{-1}}{\Sigma_{\rm p} r_{\rm p}^2/m_{\rm p}}\right)\propto r_{\rm p}^{1/2+\alpha-\beta}\,.
\label{eq:migration-timescale-lowmass}
\end{equation}
This migration rate is constant when
\begin{equation}
\alpha = \beta - \frac{1}{2}\,.
\end{equation}
For example, for a disk with $\beta=1$, exponential migration occurs only if the surface density has a slope $\alpha = 1/2$. In the following sections, the two main sources that contribute to the total torque are presented, the so-called differential Lindblad torque and the corotation torque. 

\subsubsection{Differential Lindblad torque}
\label{sec:lindblad-torque}

\begin{figure}
\centering
\includegraphics[width=\linewidth]{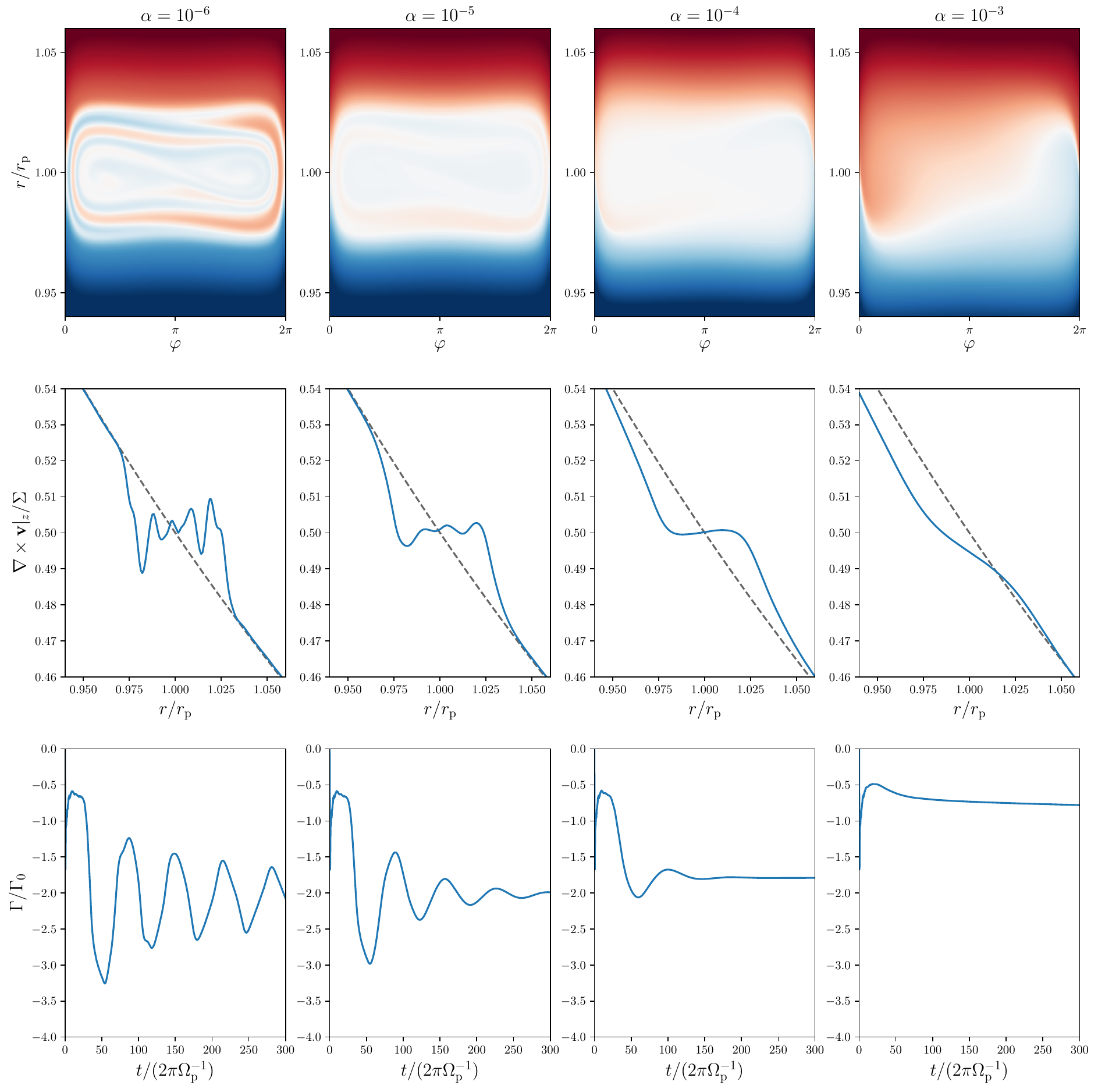}
\caption{Numerical simulations of a planet of mass $m_{\rm p} = 3\times 10^{-5} M_\star$ embedded in a global isothermal disk with an initially steep vortensity gradient ($\Sigma$ constant).
	The first row shows vortensity snapshots after a few libration timescales, at $t=180$ orbits of the planet.
	From left to right, the viscosity increases by factors of ten, from $\alpha_{\nu}=10^{-6}$ to $\alpha_\nu=10^{-3}$. 
	The surface density used to calculate the vortensity is normalized by its unperturbed value at $r = r_{\rm p}$. 
	The color map is linear, from $0.46$ (blue colors) to $0.54$ (red colors).
	The second row shows vortensity cuts along the curves $\varphi=\pi$ for each of the cases. To help quantify the restoring effect of viscosity, the initial vortensity profile it shown with dashed lines.
	The third row shows the normalized total torque experienced by the planet for each viscosity considered.
	In these panels, the effect of saturation is seen as a progressive reduction of the corotation torque. This occurs because the vortensity gradient flattens within the horseshoes, which happens by the natural motion of the material trapped in the libration region. This can be prevented by diffusion (viscosity) that is large enough to maintain the gradients active. After comparing the three rows, it is clear that the more flat the vortensity map, the weaker the corotation torque. Therefore, large negative (differential Lindblad) torques are obtained.
	In the first three columns, viscosity is low enough to allow (partial) saturation of the corotation torque, producing similar asymptotic torques that are due predominantly to the differential Lindblad torque. 
	The last column corresponds to a case for which the saturation level is very low, so the horseshoe drag can compensate for a large fraction of the differential Lindblad torque.
}
\label{fig:saturation}
\end{figure}

For each azimuthal mode $m$, one can calculate the net torque at the resonance \citep[][]{1979ApJ...233..857G} and then sum them all to obtain the so-called differential Lindblad torque.
The word differential emphasizes that this is the resulting torque from considering both the torque at the inner and outer Lindblad resonances. These torques are opposite in sign and scale linearly with the local surface density. The inner Lindblad resonance pushes the planet outward, while the outer Lindblad resonance pushes it inward.
One would argue that because the inner and outer Lindblad torques scale with the surface density of the disk, the sign of the differential Lindblad torque should depend on the steepness of the surface density gradient. This is, however, not the case for disks that are described by power-laws. The location of the Lindblad resonances is influenced by the disk pressure gradient. Larger surface density gradients lead to larger pressure gradients and reduce the disk azimuthal velocity. This, in turn, shifts both inner and outer Lindblad resonances, effectively counterbalancing the relative change in density, resulting in a differential Lindblad torque that is relatively insensitive to the surface density slope. This effect is known as pressure buffer \citep{1997Icar..126..261W}.
Not every disk is, however, well modeled by a universal power law, in which case the differential Lindblad torque could change, both in sign and magnitude. 

For typical protoplanetary disk models, where the surface density and temperature are power-laws with the disk radius, the torques at the inner and outer Lindblad resonances are asymmetric, leading to a non-negligible negative differential Lindblad torque, equal to
\begin{equation}
\Gamma_{\rm Lindblad} = \psi_{\rm L}\,\Gamma_{\rm p}\,.
\label{eq:differential-lindblad-torque}
\end{equation}
The most updated calculation of $\psi$ was done by \citet{2024ApJ...968...28T}, and it given by 
\begin{equation}
\psi_{\rm L} \simeq -\left(2.382 - 0.094\alpha + 1.297\beta\right)\,.
\label{eq:lindblad-factor}
\end{equation}
This expression is valid for three-dimensional locally isothermal disks, and it is consistent with torque measurements from numerical simulations by \citet{2017MNRAS.471.4917J}.
In the case of two-dimensional globally isothermal disks, $\psi_L = -\left(3.2 + 1.468\alpha\right)$ \citep{2002ApJ...565.1257T}.

For disks modeled as power-law disks, the differential Lindblad torque can cause Earth-sized planetary cores to migrate rapidly inward toward the central star, potentially within the lifetime of the disk.
Whether or not this is the fate of forming planets depends on several factors that set the scale of $\Gamma_{\rm p}$, such as the planet's initial radius, its growth rate, and the local disk mass.

If the differential Lindblad torque were the only torque present in the disk, planets would likely not be able to form. However, there is an additional source of torque in the disk arising from material that is closer to the planet in the corotation region, known as the corotation torque.

\subsubsection{Corotation torques}
\label{sec:corotation-torques}

The corotation torque is the torque exerted by the corotation region of the disk. In linear theory, this region corresponds to the corotation resonances discussed above (see Eq.\,\ref{eq:resonances}). In the non-linear regime, this corresponds to the horseshoe region. 
When the disk viscosity is low, linear calculations are only valid for a short time after the planet is inserted in an unperturbed (sub)Keplerian disk \citep{2009MNRAS.394.2283P}. Here, short means a $\sim$ few planetary orbits, a time much shorter than the libration timescale (see Eq.\,\ref{eq:horseshoe-libration-timescale}). This torque is called the linear corotation torque, and it was first derived by \cite{1979ApJ...233..857G}.

For each $m$-mode, the linear corotation torque is
\begin{equation}
\Gamma_{m}^{\rm C} = \frac{\pi^2 m \phi_m^2}{d\Omega/dr} \frac{d}{d r} \left(\frac{1}{\omega}\right)\,,
\label{eq:linear-corotation-torque}
\end{equation}
where $\omega =  \nabla \times v|_z/\Sigma$ is the ratio of the $z$-component of the gas vorticity to the disk surface density, known as the disk vortensity (see also Section \ref{sec:vortices}). Eq.\,\eqref{eq:linear-corotation-torque} must be evaluated at the corotation resonance radius, and shows that the larger the gradients of $\omega$, the larger the corotation torque.
While the absolute value of the corotation torque obtained from Eq.\,\eqref{eq:linear-corotation-torque} is not accurate (the actual value obtained from numerical calculations is smaller), it explains the crucial effect that a vortensity gradient has on the strength of the corotation torque.
\cite{2002ApJ...565.1257T} found that the linear corotation torque calculated through Eq.\,\eqref{eq:linear-corotation-torque} agrees well with numerical calculations if one replaces $\phi_m$ in Eq.\,\eqref{eq:linear-corotation-torque} by the effective potential that includes the disturbing potential and the enthalpy perturbation.
For a three-dimensional locally isothermal disk, after summing up all the modes, the linear corotation torque felt by a planet can be written as \citep{2024ApJ...968...28T}
\begin{equation}
\Gamma_{\rm Corrotation} = \psi_{\rm C} \Gamma_{\rm p}\,,
\label{eq:corotation-torque}
\end{equation}
where
\begin{equation}
\psi_{\rm C} = 0.63\left(\frac{3}{2}-\alpha \right) + 0.858 \beta\,.
\label{eq:psi_C}
\end{equation}
In the case of two-dimensional globally isothermal disk, \cite{2002ApJ...565.1257T} obtained  $\psi_{\rm C} \simeq 1.36\left(3/2 - \alpha \right)$.
The factor $(3/2-\alpha)$ shows that the linear corotation torque vanishes for disks in which the surface density varies as a power-law with exponent $-3/2$. This is because, for axisymmetric power-law disks, the vortensity is
\begin{equation}
\omega \equiv \frac{1}{r\Sigma}\frac{\partial}{\partial r} \left(r^2 \Omega \right) = \frac{1}{2} \frac{\Omega}{\Sigma}\,.
\end{equation}
The vortensity is a constant function as long as $\Sigma\propto\Omega\propto r^{-3/2}$.
Therefore, the value $\alpha = 3/2$ corresponds to the situation in which the disk vortensity is flat and, from Eq.\,\eqref{eq:linear-corotation-torque}, no corotation torque exists.

Once the linear phase is no longer valid, the horseshoe dynamics play a significant role in exerting the so-called horseshoe drag \citep{1991LPI....22.1463W}. This is a torque exerted by the material orbiting within the horseshoe region and performing U-turn dynamics in front and behind the planet (see Section \ref{sec:horseshoe}). 
Gas material that is in front of the planet exerts a positive torque as it is forced to a lower orbit. Conversely, the material behind the planet exerts a negative torque while being pulled outwards. These two torques cancel out only if the vortensity gradient within the horseshoe region is zero, similar to the linear corotation torque.
In two-dimensional barotropic disks, the horse-shoe drag is 
\begin{equation}
\Gamma_{\rm HS} = \frac{3}{4} \left(\frac{3}{2} -\alpha\right) \Sigma_{\rm p} \Omega_{\rm p}^2 x_{\rm s}^4\,.
\end{equation}
As discussed previously, no horseshoe drag ensues if $\alpha=3/2$. 
At first glance, the horseshoe drag and the corotation torque (Eq.\,\ref{eq:corotation-torque}) look quite different. 
However, for low-mass planets, $x_{\rm s}\propto r_{\rm p} \sqrt{m_{\rm p}/(h M_{\star})}$ (see Eq.\,\ref{eq:horseshoe-xs}) and the horseshoe drag scales as
\begin{equation}
\Gamma_{\rm HS} \propto  \left(\frac{3}{2} -\alpha\right) \Gamma_{\rm p}\,,
\end{equation}
which is the same scaling with $\Gamma_{\rm p}$ as the linear corotation and differential Lindblad torques.

As the material performs U-turns, it transfers angular momentum with the planet. As long as the horseshoe material torques the planet positively, by angular momentum conservation it must lose a fraction of its angular momentum. If the horseshoe libration region is closed and fully isolated from the disk (i.e., no angular momentum is injected into this region), the angular momentum exchange with the planet will eventually stop.
In other words, no finite region can exert a perpetual torque unless angular momentum is injected into this region externally.
The process of reducing the torque arising from the horseshoe region is called the saturation of the corotation torque.
Saturation is prevented if the horseshoe region is not closed or angular momentum is injected into it by some external mechanism. One example is viscous torques. Actually, any process that can transport angular momentum from the disk to the horseshoe region could work as long as they are efficient enough to replenish the angular momentum lost when torquing the planet.
The saturation process can be better understood by inspecting the time evolution of the vortensity within the horseshoe orbit.
Fig.\,\ref{fig:saturation} shows the results of numerical experiments for which a fixed planetary mass is considered and the disk viscosity level is gradually increased. For low-viscosity cases, the vortensity gradient flattens as the horseshoe material librates. When the viscosity is low enough, corotation torques behave like a damped oscillator, with a characteristic frequency of the order of the libration timescale.
Saturation occurs when the vortensity gradient becomes negligible within the horseshoe, which makes the total torque more negative as no corotation torque can compensate for the differential Lindblad torque.

For adiabatic disks, where the energy density evolves self-consistently, it is found that apart from the vortensity gradient, another important quantity for the corotation torque is the entropy gradient \citep[see e.g.,][]{ 2008ApJ...672.1054B}. Horseshoe motion also distributes the disk entropy similarly to what is done with vortensity. Therefore, for corotation torques, viscosity is to the vortensity as thermal diffusion is to the entropy.
In practical terms, any physical process that can affect any of the vortensity or entropy gradients within the horseshoe and keep them far from saturation can become relevant for planet migration as it could change dramatically the magnitude and even the sign of migration.

Due to the sensitivity of corotation torques to disk physics and the possibility that they can be larger than Lindblad torques, paths followed by the planets while they form depend strongly on the nature of the corotation torques. Therefore, for planet migration, it seems more reasonable to assume that the local disk structure dictates how low-mass planets migrate.

\subsection{Corotation torque of migrating planets}
\label{sec:migrating-planets}

So far, the role of migration in the development of the torques has been neglected.
Assume the situation in which a planet migrates by the action of both the differential Lindblad torque and corotation torque. 
In a frame that migrates with the planet, the local gas is seen to drift in an opposite direction. For example, if the planet migrates inwards, the planet sees the gas moving outwards through an additional radial velocity, which is an apparent effect due to its migration. In a local frame and after neglecting the curvature, the effect described is equivalent to a Galilean transformation.
This relative drift between the planet and the disk makes the horseshoes differ from the case of a fixed planet (an example of modified horseshoes due to relative drift can be seen in section \ref{sec:solids}, Fig.\,\ref{fig:dust-torque}).
As such, a migrating planet is prone to experience an additional torque due to this effect.

When a planet migrates sufficiently slow at a steady rate (i.e., the libration time is shorter than the migration time across a distance $x_{\rm s}$), \cite{2003ApJ...588..494M} have shown that the corotation torque has an additional contribution arising from relative motion, given by
\begin{equation}
\Gamma_{\rm C}^{\rm mig} = m_{\rm d} \frac{\Omega_{\rm p} }{2} r_{\rm p} \frac{d r_{\rm p}}{dt}\,,
\label{eq:mig-torque}
\end{equation}
where $m_{\rm d}$ is the vorticity weighted coorbital mass deficit, which can be written as
\begin{equation}    
m_{\rm d} = \frac{2\pi r_{\rm p} \Omega_{\rm p} x_{\rm s}}{\omega(r_{\rm p}-x_{\rm s})} \left(1 - \frac{\omega(r_{\rm p}-x_{\rm s})}{x_{\rm s}}\int_{-x_{\rm s}}^0 \frac{dx}{\omega(r_{\rm p}+x)}\right)\,.
\label{eq:mass_d}
\end{equation}
Eq.\,\eqref{eq:mass_d} shows that a torque can arise either by changes in the mass distribution within the horseshoes or by changes in its vorticity, or both.
As in the non-migrating case, the relevant quantity defining the value of the additional torque, apart from the migration rate, is the vortensity distribution. 

From Eq.\,\eqref{eq:dadt2}, the planet migration rate can be written as
\begin{equation}
\frac{dr_{\rm p}}{dt}= 2\frac{\Gamma^{\rm nmig} + \Gamma^{\rm mig}_{\rm C}}{r_{\rm p} \Omega_{\rm p}}\,,
\label{eq:rdot_mig}
\end{equation}
where the term $\Gamma^{\rm nmig}$ is the total torque (i.e., the sum of Lindblad and corotation torques) felt by the planet in the non-migrating case.
After using Eq.\,\eqref{eq:mig-torque}, the planet's migration rate is
\begin{equation}
\frac{dr_{\rm p}}{dt} = \frac{2\Gamma^{\rm nmig}}{\Delta m  r_{\rm p} \Omega_{\rm p}}\,,
\end{equation}
with $\Delta m = m_{\rm p} - m_{\rm d}$. 
Therefore, the planet migrates due to the non-migrating torques alone as if it had the effective mass $\Delta m$.
In the case $m_d \sim m_{\rm p}$, $\Delta m$ becomes small and fast migration rates can be obtained. 
However, the migration rate cannot be arbitrarily large. Eventually, the migration time over a horseshoe semi-width would become comparable to the horseshoe libration time, making it impossible to maintain a horseshoe motion. In this situation, the additional corotation torque would be reduced and the migration rate would decrease. 

Modifications of the corotation torques due to planet migration have been discovered in the context of planets that open a partial gap, for which the mass deficit $m_{\rm d}$ is triggered mainly by changes in surface density rather than by the gas vorticity \citep{2003ApJ...588..494M}. Rapid migration obtained in this regime is usually called type-III migration.
\cite{2014MNRAS.444.2031P} considered the case of low-mass planets, for which $x_{\rm s}$ is small enough so the disk vortensity variations can be neglected on scales $\sim x_{\rm s}$ and the vortensity upstream can be replaced by the disk unperturbed vortensity.
They also considered the case of low-viscosity disks, for which saturation allows a constant vortensity value, $\omega_{\rm C}$, within the horseshoe to be set. Under these assumptions, Eq.\,\eqref{eq:mass_d} simplifies to
\begin{equation}    
m_{\rm d} \simeq 4\pi  r_{\rm p} x_{\rm s}\Sigma_{\rm p}\left(1 - \frac{\omega_{\rm p}}{\omega_{\rm C}}\right)\,.
\label{eq:mass_d2}
\end{equation}
It is worth noticing that the additional torque discussed in this section arises not by planet migration itself but by the relative drift between the planet and the disk.

\subsection{Migration of massive planets}
\label{sec:migration-massive-planets}

Planets that are massive enough to repel part of the disk mass from their horseshoe region create a density depletion or gap in the disk which modifies substantially how they migrate.
While it has been argued in the past that massive planets should migrate locked to the viscous speed of the disk, their migration is not synchronized with the disk's viscous evolution.
This so-called type-II migration model \citep{1986ApJ...309..846L, 1997Icar..126..261W}, solely reliant on disk viscous drift, is decisively refuted by numerical experiments.
This occurs mainly because planetary gaps do not block the mass flow through the planetary orbit \citep[see e.g.,][]{2014ApJ...792L..10D, 2015A&A...574A..52D}.

Migration is slowed because of the reduced density in the planet's orbital radius. 
In this mass regime and for two-dimensional viscous disks that are relaxed towards the initial power-law disk far away from the planet, \cite{2018ApJ...861..140K} found that the torque felt by the planet approximately follows (within some moderate scatter) the formula
\begin{equation}
\Gamma = C \frac{\Sigma_{\rm gap}}{\Sigma_0} \Gamma_0 = \frac{C}{1+sK} \Gamma_0\,,
\label{eq:torque_massive}
\end{equation}
where all the quantities are evaluated at the planet semi-major axis, $C$ is a non-universal $\mathcal{O}(1)$ fitting constant, and $\Gamma_0$ is given by Eq.\,\eqref{eq:gamma0}. This formula is consistent with the torque felt by low-mass planets but instead of using the unperturbed density, $\Sigma_0$, the relevant quantity is the density at the bottom of the gap, $\Sigma_{\rm gap}$ (see Section \ref{sec:migration-low-mass-planets}).
Eqs.\,\eqref{eq:dadt2} and \eqref{eq:torque_massive} lead to a migration rate given by
\begin{equation}
\frac{da}{dt} = 2 C h^{-2} a \Omega_{\rm K} \left(\frac{m_{\rm p}}{M_\star}\right)^2 \left(\frac{\Sigma_{0} a^2}{m_{\rm p}}\right) \frac{1}{1+s K}\,.
\label{eq:migration_rate_massive}
\end{equation}
In the limit $sK \ll 1$, Eq.\,\eqref{eq:migration_rate_massive} shows that the migration rate scales similarly to the migration of low-mass planets. However, in the limit $sK\gg 1$ (deep gap), the migration rate scales as
\begin{equation}
\frac{da}{dt} \propto h^3 \alpha_{\nu}\, a \Omega_{\rm K} \left(\frac{\Sigma_{0} a^2}{m_{\rm p}}\right)\, = h \left(\frac{\nu}{a}\right) \left(\frac{\Sigma_{0} a^2}{m_{\rm p}}\right)\,,
\end{equation}
which shows that the migration rate is proportional to the viscous speed, $\nu/r$, and the relative disk mass ($\Sigma_0a^2/m_{\rm p}$).
Massive planets generally migrate inward more slowly than their lower-mass counterparts, yet this process can still prevent planets from maintaining distant orbits within planetary systems. 
One possibility to overcome the radial migration of massive planets is that small-mass planets, the progenitor or cores of massive planets, experience outward migration over large distances due to torques arising at corotation (see Sections \ref{sec:solids}), allowing giant planets to grow at later stages sufficiently far from the star.

Despite all the progress on the subject of migration of massive planets, a successful explanation of planet migration of massive planets from first principles is still beyond reach, mainly due to the highly non-linear nature of the problem.

\subsection{Contribution of solids to planet migration}
\label{sec:solids}

\begin{figure}
\centering
\includegraphics[width=\linewidth]{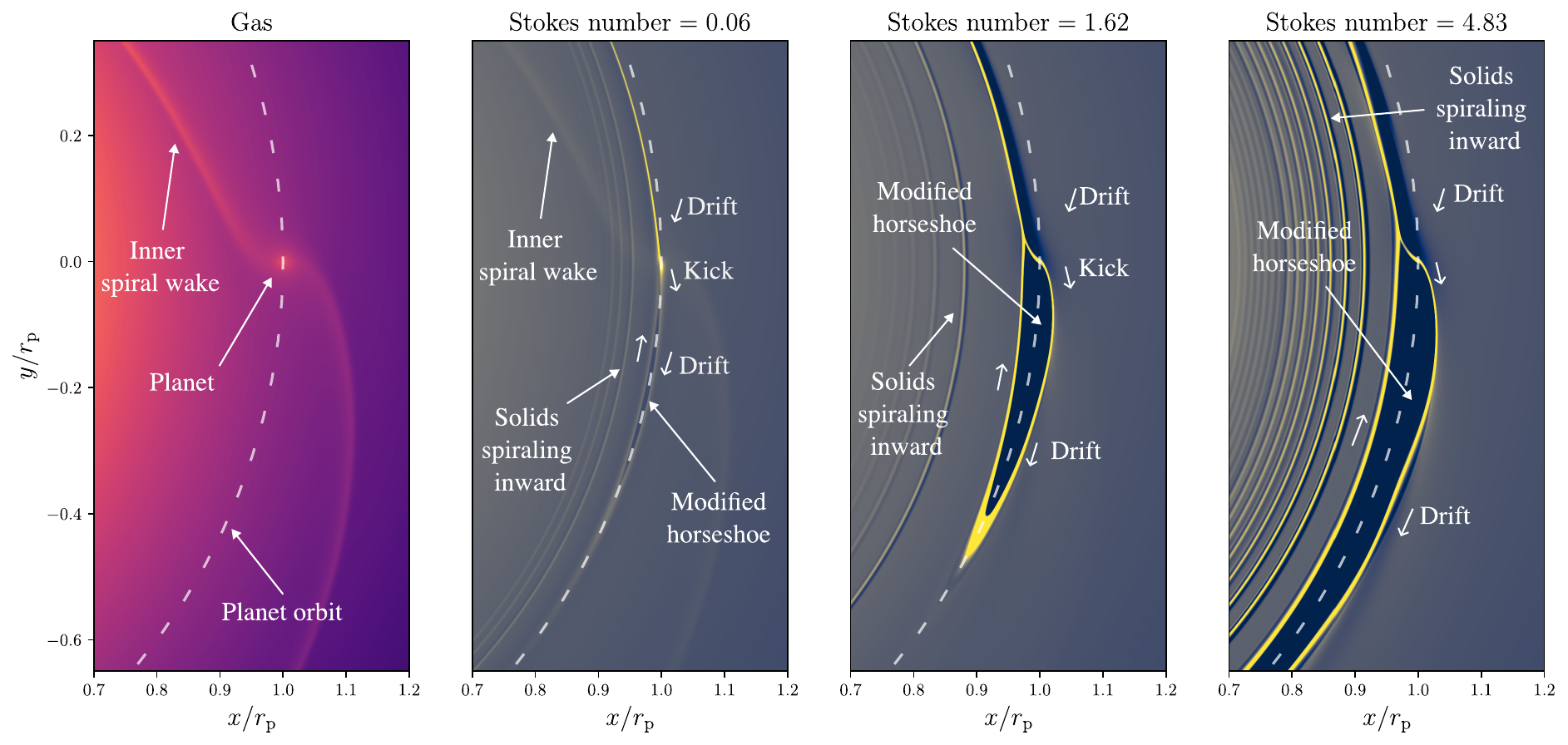}
\caption{Numerical simulation of a low-mass planet ($m_{\rm p} = 4.5 \times 10^{-6} M_\star$) and three different solid phases.
	The horseshoe motion of large particles is much wider than that of the gas. Also, the larger the particles (or Stokes numbers), the wider the partial gap. 
	Partial gap formation is explained by a gravitational kick exerted by the planet as the solid particles approach it. Also, the material initially located in the depleted region is unstable and spirals outward, leaving a depleted region of solids behind. 
	Note the differences between the gaseous phase and the solid phase. In the first, gravity and pressure are important for gas dynamics. In the second, gravity is the dominant force for large Stokes numbers. For small Stokes numbers, the gaseous spiral wake is also seen. This is because those particles tend to follow more closely the gas dynamics. In all the cases, an over-dense stream drifting inwards develops. This stream corresponds to solid concentration induced by the planet that drifts inwards by radial drift. 
	Simulation data from \cite{2018ApJ...855L..28B}.}
\label{fig:dust-torque}
\end{figure}

The initial fraction of solids in the protoplanetary disk is expected to be $\sim$ 1\% of the gaseous mass, which is the fraction measured in the interstellar medium. However, the density relative to the gas can change locally in the disk due to several factors. Solids exchange angular momentum with the gas, and in the process, they are transported vertically and radially at different velocities depending on their typical size. Particles are also subject to mutual collisions and erosion. Therefore, the particle size, shape, composition, and relative fraction can change locally over the lifetime of the disk.

For any given particle in the disk, its equation of motion is
\begin{equation}
\frac{d{\bf v}_{\rm solid}}{dt} = -\nabla{\phi} - \frac{\Omega_{\rm K}}{S_{\rm t}}\left( {\bf v}_{\rm solid} - {\bf v}\right)\,,
\label{eq:solid-drag}
\end{equation}
where $S_{\rm t}$ is the dimensionless collision rate known as the Stokes number, which measures the degree of coupling between solids and gas. This parameter contains the physical modeling of the drag force experienced by the particle. As such, the Stokes number depends on the particle geometry/density and it is also a function of the disk's local density and temperature. 
A very small Stokes number means very small particles or high gas density. In this case, particles are perfectly coupled to the gas, so they would move attached to the gas flow. The situation of a very large Stokes number implies large particles or low gas density regions, where particles are almost fully decoupled from the gas dynamics. 
It is worth mentioning that when considering solids, an additional drag force must be included in Eq.\,\eqref{eq:momentum} to allow momentum conservation. This term is typically neglected which is justified by the fact that the amount of solids is small. However, this has to be done with care as solids can easily accumulate locally, violating the working hypothesis.

Assume that, initially, solids are orbiting the central star at the Keplerian speed on a circular orbit. 
The gas does the same but at a reduced speed because of its additional positive pressure gradient that compensates part of the gravitational potential of the star. Therefore, solids embedded in the disk experience a headwind that arises from the sub-Keplerian motion of the gaseous phase (second term of Eq.\,\ref{eq:solid-drag}).
This headwind produces a negative torque on the solids that makes them drift inwards at a rate that depends on the Stokes Number. 
If the viscous drift is neglected, the gas velocity that is in equilibrium with the potential and pressure gradient is
\begin{equation}
{\bf v} \simeq r \Omega_{\rm K} \left( 1 - \frac{\eta}{2}\right) \hat{\varphi}\,.
\label{eq:gas-subkeplerian}
\end{equation}
with $\eta=h^2/2 d\log P/d\log r$. After inserting Eq.\eqref{eq:gas-subkeplerian} into Eq.\,\eqref{eq:solid-drag}, the solids radial drift is
\begin{equation}
v_{r,\rm solids} = \left(\frac{2 S_{\rm t}}{1 + S_{\rm t}^2}\right) \eta v_{\rm K}\,.
\label{eq:solids-drift}
\end{equation}
Eq.\,\eqref{eq:solids-drift} shows that particles with $S_{\rm t}<1$ exhibit a slow inward radial drift. The drift velocity reaches a maximum at $S_{\rm t} = 1$, and then decreases again for $S_{\rm t}<1$. In light of the discussion of Section \ref{sec:migrating-planets}, a modified horseshoe region is expected for significant radial drift, defined as when solids can cross the horseshoe region in a time shorter than the libration period. Since now there are two species involved, not necessarily the horseshoe motion of gas coincides with the horseshoe motion of solids.
One interesting feature that arises from the interaction of solids with a planet is that, for a given radial drift speed, there are two distinct regimes. (i) When the Stokes number is very small, the second term in Eq.\,\eqref{eq:solid-drag} dominates. Therefore, the orbits of solids are almost identical to the gaseous orbits. In this regime, the horseshoe region of the solid phase follow the gaseous phase which, for low-mass planets, has a semi-width that scales $\propto (m_{\rm p})^{1/2}$ (see Eq.\,\ref{eq:horseshoe-xs}). (ii) In the limit of large Stokes number, the first term of Eq.\,\eqref{eq:solid-drag} dominates over the second. In this regime, the equation of motion of solids is identical to that of a test particle in the restricted three-body problem. Therefore, the horseshoe semi-width scales as $\propto (m_{\rm p})^{1/3}$. In general, for low-mass planets and large Stokes number, horseshoes are wider than the ones of small Stokes number particles.

\cite{2018ApJ...855L..28B} have shown that the interaction between drifting solids and a low-mass planet leads to an asymmetric density distribution of solids close to the planet that can exert a non-negligible torque. This additional torque is positive for a wide range of Stokes numbers and planet masses and, for typical dust-to-gas mass ratios of 1\%, it can counteract the differential Lindblad torque and even make the planet migrate outwards. In addition, they have also identified the different regimes mentioned above, and have called them the gas- and gravity-dominated regimes. Interestingly, given that the torque arising from solids scales with the metallicity of the system, it should be more efficient in systems with higher metallicity, enhancing the likelihood of survival of planets and increasing the occurrence of distant giant planets in higher-metallicity stellar systems.
Fig. \ref{fig:dust-torque} shows the effect of the combined radial drift and planet torque for particles with different Stokes numbers. As long as the Stokes number increases, the modified horseshoe region in the solid phases expands. A sub-dense or partial gap forms in the solid component behind the planet, promoting the development of a positive torque that can counteract the gaseous negative torque.

As long as solids are included, it becomes natural to consider the effect that solid accretion would have on the solid dynamics and the net torque felt by the planet. \cite{2024arXiv240806076C} show that the inclusion of accretion of solids enhances the asymmetry shown in Fig.\,\ref{fig:dust-torque}, particularly for small particles, resulting in stronger positive torques for a wider region of the parameter space. This positive torque can easily promote outward migration of low-mass planets \citep[see also][]{2020MNRAS.497.5540R}.

Solid accretion onto a low-mass planet transforms the solids' gravitational energy into heat. When this heat is released by the planet, the surrounding gas gets warmer. In the simplest case, the heat would emerge from the planet isotropically, distributing the heat in the surrounding gas spherically symmetric. However, the Keplerian shear redistributes the heat within the planet's orbit.
If the disk shear is symmetric to the planet's orbit, two warm lobes (sub-dense regions) on each side of the planet would form. Since the lobes are symmetrically distributed, no net torque can be exerted by them.
However, due to the disk sub-Keplerian motion, the corotation radius (i.e., the radius that is at rest with the planet) is closer to the star than the planet. Therefore, the diffused heat would be sheared azimuthally and deposited predominantly behind the planet. Therefore, that portion of the disk that is behind the planet would be less dense than the equivalent portion in front. Therefore, the net force (and torque) exerted by the lobes would be positive. This effect is known as the Heating torque \citep{2015Natur.520...63B} and scales with the luminosity of the growing planet. A similar effect, but with an opposite sign, arises when considering the effect of thermal diffusion without any planet luminosity \citep{2014MNRAS.440..683L}.

\subsection{Migration of multiple planets}
\label{sec:multiple-planets}

\begin{figure}
\centering
\includegraphics[width=\linewidth]{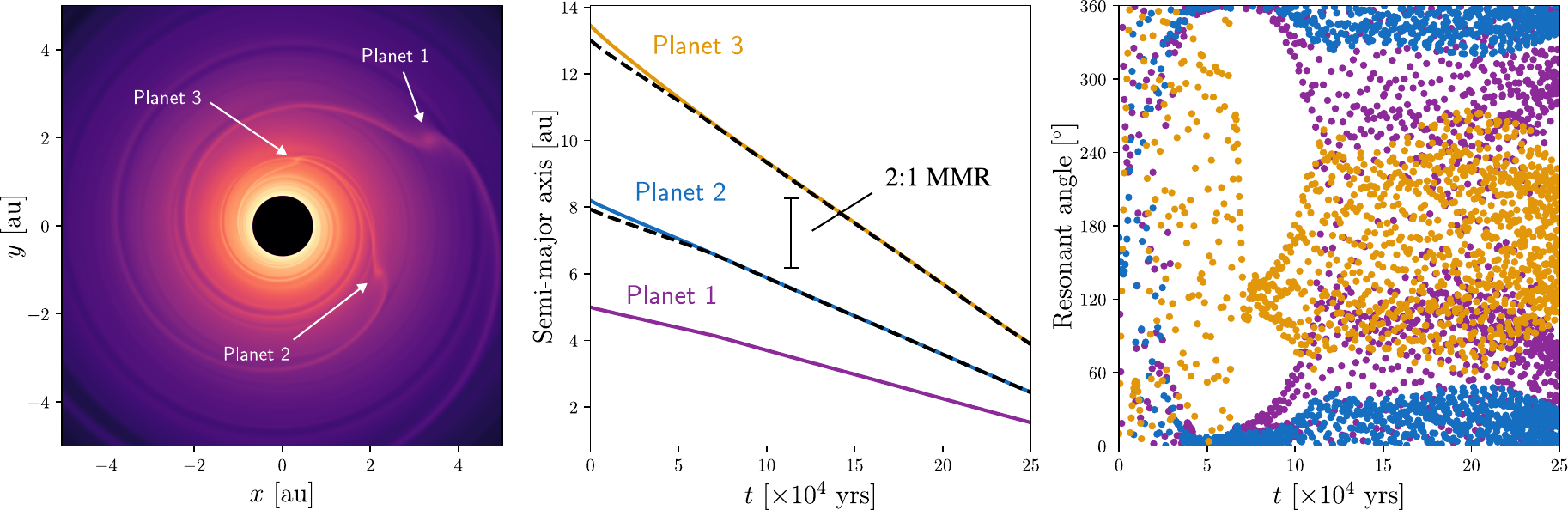}
\caption{Numerical simulation of a three-planet system.
	Masses are $m_1 = 3\times10^{-5} M_\star$, $m_2 = 6\times10^{-5} M_\star$, $m_3 = 12\times10^{-5} M_\star$. 
	All planets have an initial eccentricity $e=0.05$.
	Disk parameters are $\Sigma_0 = 400$ gr/cm$^2$, $\beta=1/2$, $\alpha = 1.0$, $h = 0.05$ and $\alpha_{\nu} = 4\times 10^{-3}$.
	The left panel shows the disk surface density at $t = 25\times 10^4$ years, where the spiral wakes of each planet can be clearly seen.
	The middle panel shows the time evolution of the semi-major axis for each planet. Dashed lines depict the relative 2:1 mean motion resonance between planets 1, 2, and 3. Immediately after the planets are inserted in the disk, they experience inward migration. Given that planet 3 is more massive than planet 2, it migrates faster, allowing convergent migration with planet 2. The same occurs between planets 2 and 1. 
	In this simulation, all planets are captured in successive 2:1 mean motion resonances, leading to a complex three-body resonance that is preserved throughout the simulation time. 
	Resonant dynamics is analyzed through the resonant angles shown in the right panel. Blue and purple colors show the resonant angles for the inner pair and outer pair, respectively. The three-body Laplace resonant angle (\citep[see e.g.,][]{1999ssd..book.....M}) is shown with orange color. Blue and purple resonant angles librate around $\sim 0^\circ$, while the orange one librates around $\sim 180^\circ$. The libration observed in all these angles is an indicator of the existence of resonance. Simulation data from \cite{2016ApJ...826...13B}.
}
\label{fig:multiple-planets}
\end{figure}

Previous discussions have primarily centered on single, isolated planets. However, planets typically form in groups within the protoplanetary disk, and their mutual gravitational perturbation could not be negligible (depending on their relative masses and mutual distances), apart from the complexity arising from additional perturbations to the disk. Evidence for this comes from the numerous pairs of exoplanets discovered in mean-motion resonances. Two planets are in mean-motion resonance when their orbital periods are a simple ratio \citep[see e.g.,][]{1999ssd..book.....M}. For instance, a 2:1 mean motion resonance happens when an inner planet orbits its star twice for every single orbit of an outer planet.
Achieving such a configuration requires dissipation to decrease the planets' relative distance and dampen their eccentricities/inclination just enough. Migration by planet-disk interaction is an elegant mechanism to explain the formation of resonant pairs or the so-called resonant chains, where several planets are in resonances all together (one example of a resonant chain is the three-body mean motion resonances of Jupiter's Galilean moons).
However, not all the observed exoplanetary systems have planets in mean motion resonances. This fact may be informing us about the efficiency of planet migration or, perhaps, that our protoplanetary disk models are still far from being realistic enough. Another possibility is that, apart from migration, additional dynamical effects are relevant to shaping the ultimate architecture of planetary systems \cite[see e.g.,][]{2014AJ....147...32G}.

To capture two (or more) planets in a mean motion resonance, they need to experience convergent migration. That is when, as the planet migrates through the disk, their relative mutual distances decreases. In this situation, they will, eventually, reach a resonance that could force them to migrate while maintaining this configuration. Convergent migration occurs if planets migrate inwards or outwards due to any of the effects discussed in Section \ref{sec:solids}. The equilibrium configuration depends on how fast planets traverse the resonance. As expected, to be trapped at a specific resonance, the migration timescale must be typically larger than the resonance libration timescale. This is, the planets relative migration must be sufficiently slow to allow them to exchange enough energy and angular momentum to maintain a stable configuration.

Fig.\,\ref{fig:multiple-planets} shows an example of planet migration for a three-planet system. Initially, the planets are located far from a resonant configuration. Given the mass distribution and initial condition, convergent migration occurs for all of the planets, resulting in a three-body resonant configuration, which is clear from the inspection of the resonant angle. The resonant configuration corresponds to a balance between planet-planet and planet-disk tidal torques.

When two or more planets are present in the disk, the gravitational dynamics and the disk response are far more complicated than those arising from a single planet. For planets that are in the linear regime, a disk response that is equal to the linear sum of the two individual responses would be a good approximation.  However, as discussed in Section \ref{sec:corotation-torques}, corotation torques are non-linear, and could be affected by the spiral wake or gravitational perturbation of the other planet. Therefore, extrapolating knowledge obtained for single planets is not straightforward. To progress on the subject, non-linear numerical calculations are crucial. However, solving multiple planets numerically is challenging. To allow the planets to migrate and interact between them, the Navier Stokes equations must be solved on a large physical domain. In addition, these studies require very long integration times to allow the resonant dynamics to operate. This problem can be circumvented by simulating only the disk region that affects the planets \citep[see e.g.,][]{2016ApJ...826...13B}.

\section{Conclusion}

Planet-disk interaction and migration are fundamental processes that shape the formation and evolution of planetary systems. As planets grow within protoplanetary disks, their gravitational influence can cause them to migrate inward or outward, significantly altering their orbital parameters.
Modeling planet-disk interactions presents a complex challenge due to the highly nonlinear equations involved. Analytical solutions are often elusive, and numerical simulations can be computationally intensive. The difficulty arises from the intricate interplay between the planets' gravitational forces and the complex disk response, governed by the equations of the hydrodynamics. Solving this response accurately is the key to characterizing planet migration.
Significant progress in our understanding of planet-disk interaction requires comprehensive numerical studies. For example, corotation torques, which are sensitive to the local disk physics near the planet's orbit, play a crucial role in migration. Therefore, improvements in disk physics modeling, particularly regarding self-consistent thermodynamics together with gas and solids dynamics, could provide valuable and more realistic insights into the migration history of growing planets.
By the end of this chapter, readers should have a firm grasp of the fundamental principles governing planet-disk interactions. This understanding will equip them to explore the topic further from theoretical, numerical, and observational perspectives.

\bigskip

\noindent {\bf Software:} All the numerical simulations presented in this chapter have been done with the publicly available code \href{https://github.com/FARGO3D/fargo3d}{FARGO3D}.

\begin{ack}[Acknowledgments] 

I am thankful to my wife, Ximena Ramos, for her support and feedback, to my daughters, Catalina and Sara, for their patience, to my colleagues, Frédéric Masset, Leonardo Krapp, Martín Pessah, Cristobal Petrovich, and to the editor, Dimitri Veras, for their feedback. The comments and suggestions of all of them have helped improve this manuscript.
I acknowledge support from ANID, QUIMAL fund ASTRO21-0039, and FONDECYT project 1231205.
\end{ack}

\seealso{Further reading on the subject: \cite{2012ARA&A..50..211K, 2014prpl.conf..667B, masset2021, 2023ASPC..534..685P} and references therein.}

\bibliographystyle{Harvard}

\begin{thebibliography*}{40}
\providecommand{\bibtype}[1]{}
\providecommand{\natexlab}[1]{#1}
{\catcode`\|=0\catcode`\#=12\catcode`\@=11\catcode`\\=12
	|immediate|write|@auxout{\expandafter\ifx\csname
		natexlab\endcsname\relax\gdef\natexlab#1{#1}\fi}}
\renewcommand{\url}[1]{{\tt #1}}
\providecommand{\urlprefix}{URL }
\expandafter\ifx\csname urlstyle\endcsname\relax
\providecommand{\doi}[1]{doi:\discretionary{}{}{}#1}\else
\providecommand{\doi}{doi:\discretionary{}{}{}\begingroup
	\urlstyle{rm}\Url}\fi
\providecommand{\bibinfo}[2]{#2}
\providecommand{\eprint}[2][]{\url{#2}}

\bibtype{Article}%
\bibitem[{Artymowicz}(1993)]{1993ApJ...419..155A}
\bibinfo{author}{{Artymowicz} P} (\bibinfo{year}{1993}), \bibinfo{month}{Dec.}
\bibinfo{title}{{On the Wave Excitation and a Generalized Torque Formula for
		Lindblad Resonances Excited by External Potential}}.
\bibinfo{journal}{{\em \apj}} \bibinfo{volume}{419}: \bibinfo{pages}{155}.
\bibinfo{doi}{\doi{10.1086/173469}}.

\bibtype{Article}%
\bibitem[{Bae} and {Zhu}(2018)]{2018ApJ...859..118B}
\bibinfo{author}{{Bae} J} and  \bibinfo{author}{{Zhu} Z}
(\bibinfo{year}{2018}), \bibinfo{month}{Jun.}
\bibinfo{title}{{Planet-driven Spiral Arms in Protoplanetary Disks. I.
		Formation Mechanism}}.
\bibinfo{journal}{{\em \apj}} \bibinfo{volume}{859} (\bibinfo{number}{2}),
\bibinfo{eid}{118}. \bibinfo{doi}{\doi{10.3847/1538-4357/aabf8c}}.
\eprint{1711.08161}.

\bibtype{Article}%
\bibitem[{Baruteau} and {Masset}(2008)]{2008ApJ...672.1054B}
\bibinfo{author}{{Baruteau} C} and  \bibinfo{author}{{Masset} F}
(\bibinfo{year}{2008}), \bibinfo{month}{Jan.}
\bibinfo{title}{{On the Corotation Torque in a Radiatively Inefficient Disk}}.
\bibinfo{journal}{{\em \apj}} \bibinfo{volume}{672} (\bibinfo{number}{2}):
\bibinfo{pages}{1054--1067}. \bibinfo{doi}{\doi{10.1086/523667}}.
\eprint{0709.2617}.

\bibtype{Inproceedings}%
\bibitem[{Baruteau} et al.(2014)]{2014prpl.conf..667B}
\bibinfo{author}{{Baruteau} C}, \bibinfo{author}{{Crida} A},
\bibinfo{author}{{Paardekooper} SJ}, \bibinfo{author}{{Masset} F},
\bibinfo{author}{{Guilet} J}, \bibinfo{author}{{Bitsch} B},
\bibinfo{author}{{Nelson} R}, \bibinfo{author}{{Kley} W} and
\bibinfo{author}{{Papaloizou} J} (\bibinfo{year}{2014}),
\bibinfo{month}{Jan.}, \bibinfo{title}{{Planet-Disk Interactions and Early
		Evolution of Planetary Systems}}, \bibinfo{editor}{{Beuther} H},
\bibinfo{editor}{{Klessen} RS}, \bibinfo{editor}{{Dullemond} CP} and
\bibinfo{editor}{{Henning} T}, (Eds.), \bibinfo{booktitle}{Protostars and
	Planets VI},  \bibinfo{pages}{667--689}, \eprint{1312.4293}.

\bibtype{Article}%
\bibitem[{Ben{\'\i}tez-Llambay} and {Pessah}(2018)]{2018ApJ...855L..28B}
\bibinfo{author}{{Ben{\'\i}tez-Llambay} P} and  \bibinfo{author}{{Pessah} ME}
(\bibinfo{year}{2018}), \bibinfo{month}{Mar.}
\bibinfo{title}{{Torques Induced by Scattered Pebble-flow in Protoplanetary
		Disks}}.
\bibinfo{journal}{{\em \apjl}} \bibinfo{volume}{855} (\bibinfo{number}{2}),
\bibinfo{eid}{L28}. \bibinfo{doi}{\doi{10.3847/2041-8213/aab2ae}}.
\eprint{1801.07913}.

\bibtype{Article}%
\bibitem[{Ben{\'\i}tez-Llambay} et al.(2015)]{2015Natur.520...63B}
\bibinfo{author}{{Ben{\'\i}tez-Llambay} P}, \bibinfo{author}{{Masset} F},
\bibinfo{author}{{Koenigsberger} G} and  \bibinfo{author}{{Szul{\'a}gyi} J}
(\bibinfo{year}{2015}), \bibinfo{month}{Apr.}
\bibinfo{title}{{Planet heating prevents inward migration of planetary cores}}.
\bibinfo{journal}{{\em \nat}} \bibinfo{volume}{520} (\bibinfo{number}{7545}):
\bibinfo{pages}{63--65}. \bibinfo{doi}{\doi{10.1038/nature14277}}.
\eprint{1510.01778}.

\bibtype{Article}%
\bibitem[{Ben{\'\i}tez-Llambay} et al.(2016)]{2016ApJ...826...13B}
\bibinfo{author}{{Ben{\'\i}tez-Llambay} P}, \bibinfo{author}{{Ramos} XS},
\bibinfo{author}{{Beaug{\'e}} C} and  \bibinfo{author}{{Masset} FS}
(\bibinfo{year}{2016}), \bibinfo{month}{Jul.}
\bibinfo{title}{{Long-term and Large-scale Hydrodynamical Simulations of
		Migrating Planets}}.
\bibinfo{journal}{{\em \apj}} \bibinfo{volume}{826} (\bibinfo{number}{1}),
\bibinfo{eid}{13}. \bibinfo{doi}{\doi{10.3847/0004-637X/826/1/13}}.
\eprint{1605.01618}.

\bibtype{Article}%
\bibitem[{Chrenko} et al.(2024)]{2024arXiv240806076C}
\bibinfo{author}{{Chrenko} O}, \bibinfo{author}{{Chametla} RO},
\bibinfo{author}{{Masset} FS}, \bibinfo{author}{{Baruteau} C} and
\bibinfo{author}{{Bro{\v{z}}} M} (\bibinfo{year}{2024}),
\bibinfo{month}{Aug.}
\bibinfo{title}{{Pebble-driven migration of low-mass planets in the 2D regime
		of pebble accretion}}.
\bibinfo{journal}{{\em arXiv e-prints}} ,
\bibinfo{eid}{arXiv:2408.06076}\bibinfo{doi}{\doi{10.48550/arXiv.2408.06076}}.
\eprint{2408.06076}.

\bibtype{Article}%
\bibitem[{Dempsey} et al.(2020)]{2020ApJ...891..108D}
\bibinfo{author}{{Dempsey} AM}, \bibinfo{author}{{Lee} WK} and
\bibinfo{author}{{Lithwick} Y} (\bibinfo{year}{2020}), \bibinfo{month}{Mar.}
\bibinfo{title}{{Pileups and Migration Rates for Planets in Low-mass Disks}}.
\bibinfo{journal}{{\em \apj}} \bibinfo{volume}{891} (\bibinfo{number}{2}),
\bibinfo{eid}{108}. \bibinfo{doi}{\doi{10.3847/1538-4357/ab723c}}.
\eprint{1908.02326}.

\bibtype{Article}%
\bibitem[{Duffell} et al.(2014)]{2014ApJ...792L..10D}
\bibinfo{author}{{Duffell} PC}, \bibinfo{author}{{Haiman} Z},
\bibinfo{author}{{MacFadyen} AI}, \bibinfo{author}{{D'Orazio} DJ} and
\bibinfo{author}{{Farris} BD} (\bibinfo{year}{2014}), \bibinfo{month}{Sep.}
\bibinfo{title}{{The Migration of Gap-opening Planets is Not Locked to Viscous
		Disk Evolution}}.
\bibinfo{journal}{{\em \apjl}} \bibinfo{volume}{792} (\bibinfo{number}{1}),
\bibinfo{eid}{L10}. \bibinfo{doi}{\doi{10.1088/2041-8205/792/1/L10}}.
\eprint{1405.3711}.

\bibtype{Article}%
\bibitem[{D{\"u}rmann} and {Kley}(2015)]{2015A&A...574A..52D}
\bibinfo{author}{{D{\"u}rmann} C} and  \bibinfo{author}{{Kley} W}
(\bibinfo{year}{2015}), \bibinfo{month}{Feb.}
\bibinfo{title}{{Migration of massive planets in accreting disks}}.
\bibinfo{journal}{{\em \aap}} \bibinfo{volume}{574}, \bibinfo{eid}{A52}.
\bibinfo{doi}{\doi{10.1051/0004-6361/201424837}}.
\eprint{1411.3190}.

\bibtype{Article}%
\bibitem[{Fung} and {Dong}(2015)]{2015ApJ...815L..21F}
\bibinfo{author}{{Fung} J} and  \bibinfo{author}{{Dong} R}
(\bibinfo{year}{2015}), \bibinfo{month}{Dec.}
\bibinfo{title}{{Inferring Planet Mass from Spiral Structures in Protoplanetary
		Disks}}.
\bibinfo{journal}{{\em \apjl}} \bibinfo{volume}{815} (\bibinfo{number}{2}),
\bibinfo{eid}{L21}. \bibinfo{doi}{\doi{10.1088/2041-8205/815/2/L21}}.
\eprint{1511.01178}.

\bibtype{Article}%
\bibitem[{Fung} et al.(2019)]{2019ApJ...887..152F}
\bibinfo{author}{{Fung} J}, \bibinfo{author}{{Zhu} Z} and
\bibinfo{author}{{Chiang} E} (\bibinfo{year}{2019}), \bibinfo{month}{Dec.}
\bibinfo{title}{{Circumplanetary Disk Dynamics in the Isothermal and Adiabatic
		Limits}}.
\bibinfo{journal}{{\em \apj}} \bibinfo{volume}{887} (\bibinfo{number}{2}),
\bibinfo{eid}{152}. \bibinfo{doi}{\doi{10.3847/1538-4357/ab53da}}.
\eprint{1909.09655}.

\bibtype{Article}%
\bibitem[{Goldreich} and {Schlichting}(2014)]{2014AJ....147...32G}
\bibinfo{author}{{Goldreich} P} and  \bibinfo{author}{{Schlichting} HE}
(\bibinfo{year}{2014}), \bibinfo{month}{Feb.}
\bibinfo{title}{{Overstable Librations can Account for the Paucity of Mean
		Motion Resonances among Exoplanet Pairs}}.
\bibinfo{journal}{{\em \aj}} \bibinfo{volume}{147} (\bibinfo{number}{2}),
\bibinfo{eid}{32}. \bibinfo{doi}{\doi{10.1088/0004-6256/147/2/32}}.
\eprint{1308.4688}.

\bibtype{Article}%
\bibitem[{Goldreich} and {Tremaine}(1979)]{1979ApJ...233..857G}
\bibinfo{author}{{Goldreich} P} and  \bibinfo{author}{{Tremaine} S}
(\bibinfo{year}{1979}), \bibinfo{month}{Nov.}
\bibinfo{title}{{The excitation of density waves at the Lindblad and corotation
		resonances by an external potential.}}
\bibinfo{journal}{{\em \apj}} \bibinfo{volume}{233}: \bibinfo{pages}{857--871}.
\bibinfo{doi}{\doi{10.1086/157448}}.

\bibtype{Article}%
\bibitem[{Goodman} and {Rafikov}(2001)]{2001ApJ...552..793G}
\bibinfo{author}{{Goodman} J} and  \bibinfo{author}{{Rafikov} RR}
(\bibinfo{year}{2001}), \bibinfo{month}{May}.
\bibinfo{title}{{Planetary Torques as the Viscosity of Protoplanetary Disks}}.
\bibinfo{journal}{{\em \apj}} \bibinfo{volume}{552} (\bibinfo{number}{2}):
\bibinfo{pages}{793--802}. \bibinfo{doi}{\doi{10.1086/320572}}.
\eprint{astro-ph/0010576}.

\bibtype{Article}%
\bibitem[{Jim{\'e}nez} and {Masset}(2017)]{2017MNRAS.471.4917J}
\bibinfo{author}{{Jim{\'e}nez} MA} and  \bibinfo{author}{{Masset} FS}
(\bibinfo{year}{2017}), \bibinfo{month}{Nov.}
\bibinfo{title}{{Improved torque formula for low- and intermediate-mass
		planetary migration}}.
\bibinfo{journal}{{\em \mnras}} \bibinfo{volume}{471} (\bibinfo{number}{4}):
\bibinfo{pages}{4917--4929}. \bibinfo{doi}{\doi{10.1093/mnras/stx1946}}.
\eprint{1707.08988}.

\bibtype{Article}%
\bibitem[{Kanagawa} et al.(2017)]{2017PASJ...69...97K}
\bibinfo{author}{{Kanagawa} KD}, \bibinfo{author}{{Tanaka} H},
\bibinfo{author}{{Muto} T} and  \bibinfo{author}{{Tanigawa} T}
(\bibinfo{year}{2017}), \bibinfo{month}{Dec.}
\bibinfo{title}{{Modelling of deep gaps created by giant planets in
		protoplanetary disks}}.
\bibinfo{journal}{{\em \pasj}} \bibinfo{volume}{69} (\bibinfo{number}{6}),
\bibinfo{eid}{97}. \bibinfo{doi}{\doi{10.1093/pasj/psx114}}.
\eprint{1609.02706}.

\bibtype{Article}%
\bibitem[{Kanagawa} et al.(2018)]{2018ApJ...861..140K}
\bibinfo{author}{{Kanagawa} KD}, \bibinfo{author}{{Tanaka} H} and
\bibinfo{author}{{Szuszkiewicz} E} (\bibinfo{year}{2018}),
\bibinfo{month}{Jul.}
\bibinfo{title}{{Radial Migration of Gap-opening Planets in Protoplanetary
		Disks. I. The Case of a Single Planet}}.
\bibinfo{journal}{{\em \apj}} \bibinfo{volume}{861} (\bibinfo{number}{2}),
\bibinfo{eid}{140}. \bibinfo{doi}{\doi{10.3847/1538-4357/aac8d9}}.
\eprint{1805.11101}.

\bibtype{Article}%
\bibitem[{Kley} and {Nelson}(2012)]{2012ARA&A..50..211K}
\bibinfo{author}{{Kley} W} and  \bibinfo{author}{{Nelson} RP}
(\bibinfo{year}{2012}), \bibinfo{month}{Sep.}
\bibinfo{title}{{Planet-Disk Interaction and Orbital Evolution}}.
\bibinfo{journal}{{\em \araa}} \bibinfo{volume}{50}: \bibinfo{pages}{211--249}.
\bibinfo{doi}{\doi{10.1146/annurev-astro-081811-125523}}.
\eprint{1203.1184}.

\bibtype{Article}%
\bibitem[{Lega} et al.(2014)]{2014MNRAS.440..683L}
\bibinfo{author}{{Lega} E}, \bibinfo{author}{{Crida} A},
\bibinfo{author}{{Bitsch} B} and  \bibinfo{author}{{Morbidelli} A}
(\bibinfo{year}{2014}), \bibinfo{month}{May}.
\bibinfo{title}{{Migration of Earth-sized planets in 3D radiative discs}}.
\bibinfo{journal}{{\em \mnras}} \bibinfo{volume}{440} (\bibinfo{number}{1}):
\bibinfo{pages}{683--695}. \bibinfo{doi}{\doi{10.1093/mnras/stu304}}.
\eprint{1402.2834}.

\bibtype{Article}%
\bibitem[{Lin} and {Papaloizou}(1986)]{1986ApJ...309..846L}
\bibinfo{author}{{Lin} DNC} and  \bibinfo{author}{{Papaloizou} J}
(\bibinfo{year}{1986}), \bibinfo{month}{Oct.}
\bibinfo{title}{{On the Tidal Interaction between Protoplanets and the
		Protoplanetary Disk. III. Orbital Migration of Protoplanets}}.
\bibinfo{journal}{{\em \apj}} \bibinfo{volume}{309}: \bibinfo{pages}{846}.
\bibinfo{doi}{\doi{10.1086/164653}}.

\bibtype{Article}%
\bibitem[{Lovelace} et al.(1999)]{1999ApJ...513..805L}
\bibinfo{author}{{Lovelace} RVE}, \bibinfo{author}{{Li} H},
\bibinfo{author}{{Colgate} SA} and  \bibinfo{author}{{Nelson} AF}
(\bibinfo{year}{1999}), \bibinfo{month}{Mar.}
\bibinfo{title}{{Rossby Wave Instability of Keplerian Accretion Disks}}.
\bibinfo{journal}{{\em \apj}} \bibinfo{volume}{513} (\bibinfo{number}{2}):
\bibinfo{pages}{805--810}. \bibinfo{doi}{\doi{10.1086/306900}}.
\eprint{astro-ph/9809321}.

\bibtype{Misc}%
\bibitem[Masset(2021)]{masset2021}
\bibinfo{author}{Masset FS} (\bibinfo{year}{2021}), \bibinfo{month}{09}.
\bibinfo{title}{Migration of low-mass planets}.
\bibinfo{doi}{\doi{10.1093/acrefore/9780190647926.013.192}}.

\bibtype{Article}%
\bibitem[{Masset} and {Ben{\'\i}tez-Llambay}(2016)]{2016ApJ...817...19M}
\bibinfo{author}{{Masset} FS} and  \bibinfo{author}{{Ben{\'\i}tez-Llambay} P}
(\bibinfo{year}{2016}), \bibinfo{month}{Jan.}
\bibinfo{title}{{Horseshoe Drag in Three-dimensional Globally Isothermal
		Disks}}.
\bibinfo{journal}{{\em \apj}} \bibinfo{volume}{817} (\bibinfo{number}{1}),
\bibinfo{eid}{19}. \bibinfo{doi}{\doi{10.3847/0004-637X/817/1/19}}.
\eprint{1511.07946}.

\bibtype{Article}%
\bibitem[{Masset} and {Papaloizou}(2003)]{2003ApJ...588..494M}
\bibinfo{author}{{Masset} FS} and  \bibinfo{author}{{Papaloizou} JCB}
(\bibinfo{year}{2003}), \bibinfo{month}{May}.
\bibinfo{title}{{Runaway Migration and the Formation of Hot Jupiters}}.
\bibinfo{journal}{{\em \apj}} \bibinfo{volume}{588} (\bibinfo{number}{1}):
\bibinfo{pages}{494--508}. \bibinfo{doi}{\doi{10.1086/373892}}.
\eprint{astro-ph/0301171}.

\bibtype{Article}%
\bibitem[{Masset} et al.(2006)]{2006ApJ...652..730M}
\bibinfo{author}{{Masset} FS}, \bibinfo{author}{{D'Angelo} G} and
\bibinfo{author}{{Kley} W} (\bibinfo{year}{2006}), \bibinfo{month}{Nov.}
\bibinfo{title}{{On the Migration of Protogiant Solid Cores}}.
\bibinfo{journal}{{\em \apj}} \bibinfo{volume}{652} (\bibinfo{number}{1}):
\bibinfo{pages}{730--745}. \bibinfo{doi}{\doi{10.1086/507515}}.
\eprint{astro-ph/0607155}.

\bibtype{Book}%
\bibitem[{Murray} and {Dermott}(1999)]{1999ssd..book.....M}
\bibinfo{author}{{Murray} CD} and  \bibinfo{author}{{Dermott} SF}
(\bibinfo{year}{1999}).
\bibinfo{title}{{Solar System Dynamics}}.
\bibinfo{doi}{\doi{10.1017/CBO9781139174817}}.

\bibtype{Article}%
\bibitem[{Nelson} et al.(2013)]{2013MNRAS.435.2610N}
\bibinfo{author}{{Nelson} RP}, \bibinfo{author}{{Gressel} O} and
\bibinfo{author}{{Umurhan} OM} (\bibinfo{year}{2013}), \bibinfo{month}{Nov.}
\bibinfo{title}{{Linear and non-linear evolution of the vertical shear
		instability in accretion discs}}.
\bibinfo{journal}{{\em \mnras}} \bibinfo{volume}{435} (\bibinfo{number}{3}):
\bibinfo{pages}{2610--2632}. \bibinfo{doi}{\doi{10.1093/mnras/stt1475}}.
\eprint{1209.2753}.

\bibtype{Article}%
\bibitem[{Ogilvie} and {Lubow}(2002)]{2002MNRAS.330..950O}
\bibinfo{author}{{Ogilvie} GI} and  \bibinfo{author}{{Lubow} SH}
(\bibinfo{year}{2002}), \bibinfo{month}{Mar.}
\bibinfo{title}{{On the wake generated by a planet in a disc}}.
\bibinfo{journal}{{\em \mnras}} \bibinfo{volume}{330} (\bibinfo{number}{4}):
\bibinfo{pages}{950--954}.
\bibinfo{doi}{\doi{10.1046/j.1365-8711.2002.05148.x}}.
\eprint{astro-ph/0111265}.

\bibtype{Article}%
\bibitem[{Paardekooper}(2014)]{2014MNRAS.444.2031P}
\bibinfo{author}{{Paardekooper} SJ} (\bibinfo{year}{2014}),
\bibinfo{month}{Nov.}
\bibinfo{title}{{Dynamical corotation torques on low-mass planets}}.
\bibinfo{journal}{{\em \mnras}} \bibinfo{volume}{444} (\bibinfo{number}{3}):
\bibinfo{pages}{2031--2042}. \bibinfo{doi}{\doi{10.1093/mnras/stu1542}}.
\eprint{1409.0372}.

\bibtype{Article}%
\bibitem[{Paardekooper} and
{Papaloizou}(2009{\natexlab{a}})]{2009MNRAS.394.2283P}
\bibinfo{author}{{Paardekooper} SJ} and  \bibinfo{author}{{Papaloizou} JCB}
(\bibinfo{year}{2009}{\natexlab{a}}), \bibinfo{month}{Apr.}
\bibinfo{title}{{On corotation torques, horseshoe drag and the possibility of
		sustained stalled or outward protoplanetary migration}}.
\bibinfo{journal}{{\em \mnras}} \bibinfo{volume}{394} (\bibinfo{number}{4}):
\bibinfo{pages}{2283--2296}.
\bibinfo{doi}{\doi{10.1111/j.1365-2966.2009.14511.x}}.
\eprint{0901.2265}.

\bibtype{Article}%
\bibitem[{Paardekooper} and
{Papaloizou}(2009{\natexlab{b}})]{2009MNRAS.394.2297P}
\bibinfo{author}{{Paardekooper} SJ} and  \bibinfo{author}{{Papaloizou} JCB}
(\bibinfo{year}{2009}{\natexlab{b}}), \bibinfo{month}{Apr.}
\bibinfo{title}{{On the width and shape of the corotation region for low-mass
		planets}}.
\bibinfo{journal}{{\em \mnras}} \bibinfo{volume}{394} (\bibinfo{number}{4}):
\bibinfo{pages}{2297--2309}.
\bibinfo{doi}{\doi{10.1111/j.1365-2966.2009.14512.x}}.
\eprint{0901.2263}.

\bibtype{Inproceedings}%
\bibitem[{Paardekooper} et al.(2023)]{2023ASPC..534..685P}
\bibinfo{author}{{Paardekooper} S}, \bibinfo{author}{{Dong} R},
\bibinfo{author}{{Duffell} P}, \bibinfo{author}{{Fung} J},
\bibinfo{author}{{Masset} FS}, \bibinfo{author}{{Ogilvie} G} and
\bibinfo{author}{{Tanaka} H} (\bibinfo{year}{2023}), \bibinfo{month}{Jul.},
\bibinfo{title}{{Planet-Disk Interactions and Orbital Evolution}},
\bibinfo{editor}{{Inutsuka} S}, \bibinfo{editor}{{Aikawa} Y},
\bibinfo{editor}{{Muto} T}, \bibinfo{editor}{{Tomida} K} and
\bibinfo{editor}{{Tamura} M}, (Eds.), \bibinfo{booktitle}{Protostars and
	Planets VII}, \bibinfo{series}{Astronomical Society of the Pacific Conference
	Series}, \bibinfo{volume}{534}, pp. \bibinfo{pages}{685},
\eprint{2203.09595}.

\bibtype{Inproceedings}%
\bibitem[{Pascucci} et al.(2023)]{2023ASPC..534..567P}
\bibinfo{author}{{Pascucci} I}, \bibinfo{author}{{Cabrit} S},
\bibinfo{author}{{Edwards} S}, \bibinfo{author}{{Gorti} U},
\bibinfo{author}{{Gressel} O} and  \bibinfo{author}{{Suzuki} TK}
(\bibinfo{year}{2023}), \bibinfo{month}{Jul.}, \bibinfo{title}{{The Role of
		Disk Winds in the Evolution and Dispersal of Protoplanetary Disks}},
\bibinfo{editor}{{Inutsuka} S}, \bibinfo{editor}{{Aikawa} Y},
\bibinfo{editor}{{Muto} T}, \bibinfo{editor}{{Tomida} K} and
\bibinfo{editor}{{Tamura} M}, (Eds.), \bibinfo{booktitle}{Protostars and
	Planets VII}, \bibinfo{series}{Astronomical Society of the Pacific Conference
	Series}, \bibinfo{volume}{534}, pp. \bibinfo{pages}{567},
\eprint{2203.10068}.

\bibtype{Article}%
\bibitem[{Reg{\'a}ly}(2020)]{2020MNRAS.497.5540R}
\bibinfo{author}{{Reg{\'a}ly} Z} (\bibinfo{year}{2020}), \bibinfo{month}{Oct.}
\bibinfo{title}{{Torques felt by solid accreting planets}}.
\bibinfo{journal}{{\em \mnras}} \bibinfo{volume}{497} (\bibinfo{number}{4}):
\bibinfo{pages}{5540--5549}. \bibinfo{doi}{\doi{10.1093/mnras/staa2181}}.
\eprint{2007.11072}.

\bibtype{Article}%
\bibitem[{Tanaka} and {Okada}(2024)]{2024ApJ...968...28T}
\bibinfo{author}{{Tanaka} H} and  \bibinfo{author}{{Okada} K}
(\bibinfo{year}{2024}), \bibinfo{month}{Jun.}
\bibinfo{title}{{Three-dimensional Interaction between a Planet and an
		Isothermal Gaseous Disk. III. Locally Isothermal Cases}}.
\bibinfo{journal}{{\em \apj}} \bibinfo{volume}{968} (\bibinfo{number}{1}),
\bibinfo{eid}{28}. \bibinfo{doi}{\doi{10.3847/1538-4357/ad410d}}.
\eprint{2404.12521}.

\bibtype{Article}%
\bibitem[{Tanaka} et al.(2002)]{2002ApJ...565.1257T}
\bibinfo{author}{{Tanaka} H}, \bibinfo{author}{{Takeuchi} T} and
\bibinfo{author}{{Ward} WR} (\bibinfo{year}{2002}), \bibinfo{month}{Feb.}
\bibinfo{title}{{Three-Dimensional Interaction between a Planet and an
		Isothermal Gaseous Disk. I. Corotation and Lindblad Torques and Planet
		Migration}}.
\bibinfo{journal}{{\em \apj}} \bibinfo{volume}{565} (\bibinfo{number}{2}):
\bibinfo{pages}{1257--1274}. \bibinfo{doi}{\doi{10.1086/324713}}.

\bibtype{Inproceedings}%
\bibitem[{Ward}(1991)]{1991LPI....22.1463W}
\bibinfo{author}{{Ward} WR} (\bibinfo{year}{1991}), \bibinfo{month}{Mar.},
\bibinfo{title}{{Horsehoe Orbit Drag}}, \bibinfo{booktitle}{Lunar and
	Planetary Science Conference}, \bibinfo{series}{Lunar and Planetary Science
	Conference}, \bibinfo{volume}{22}, pp. \bibinfo{pages}{1463}.

\bibtype{Article}%
\bibitem[{Ward}(1997)]{1997Icar..126..261W}
\bibinfo{author}{{Ward} WR} (\bibinfo{year}{1997}), \bibinfo{month}{Apr.}
\bibinfo{title}{{Protoplanet Migration by Nebula Tides}}.
\bibinfo{journal}{{\em \icarus}} \bibinfo{volume}{126} (\bibinfo{number}{2}):
\bibinfo{pages}{261--281}. \bibinfo{doi}{\doi{10.1006/icar.1996.5647}}.

\end{thebibliography*}

\end{document}